\documentclass[reprint,reprint,amsmath,superscriptaddress,amssymb,aps,nofootinbib,pra]{revtex4-2}
\usepackage{graphicx}   % Include figure files
\usepackage{dcolumn}    % Align table columns on decimal
\usepackage{bm}         % bold math
\usepackage{hyperref}   % add hypertext capabilities
\usepackage{float}
\usepackage{soul}
\usepackage{endnotes}
\begin{document}

\preprint{APS/123-QED}

% Commands for the review
% \newcommand{\rev}[2]{\textcolor{red}{#1}
%  \footnote{#2}}
% % \let \checknote \endnote
% \newcommand{\hil}[2]{\colorbox{LimeGreen}{#1}\footnote{#2}}
% \newcommand{\hilye}[2]{\colorbox{Goldenrod}{#1}\footnote{#2}}
% % \newcommand{\mysectodo}[1]{\addcontentsline{tdo}{todo}{#1}}

% % \newcommand{\tdn}[2] {\stepcounter{todocounter}\sethlcolor{Goldenrod}\hl{#1
% % $^{\thetodocounter}$}\mysectodo{\thetodocounter : #2}}

% \newcommand{\tdn}[2]{\sethlcolor{Goldenrod}\hl{#1}\endnote{#2}}

% \newcommand{\tocite}{\textcolor{red}{[CITE]}}

% \renewcommand*{\notesname}{Notes}
% % \makeatletter
% \renewcommand*{\enoteheading}{\subsection*{\large{\textbf{\notesname}}}}%
% \newcommand{\entoc}{\theendnotes}
% Commands for the review%%%

\title{An analog-electronic implementation of a harmonic oscillator recurrent neural network}

\author{Pedro Carvalho}
% \email{prfdecarvalho@gmail.com}
\email{pedro.carvalho@esi-frankfurt.de}
\affiliation{Ernst Strüngmann Institute, Frankfurt am Main, Germany}%

\author{Bernd Ulmann}%
\email{ulmann@anabrid.de}
\affiliation{anabrid GmbH; FOM University of Applied Sciences, Frankfurt am
Main, Germany}%

\author{Wolf Singer}%
\email{wolf.singer@brain.mpg.de}
\affiliation{Ernst Strüngmann Institute, Frankfurt am Main, Germany}%
\affiliation{Max Planck Institute for Brain Research, Frankfurt am Main,
Germany}%

\author{Felix Effenberger}%
\email{felix.effenberger@esi-frankfurt.de}%
\affiliation{Ernst Strüngmann Institute, Frankfurt am Main, Germany}%

\date{\today}

\begin{abstract}
Oscillatory recurrent networks, such as the Harmonic Oscillator Recurrent Network (HORN) model, offer advantages in parameter efficiency, learning speed, and robustness relative to traditional non-oscillating architectures.
Yet, while many implementations of physical neural networks exploiting attractor dynamics have been studied, implementations of oscillatory models in analog-electronic hardware that utilize the networks' transient dynamics so far are lacking.
This study explores the feasibility of implementing HORNs in analog-electronic hardware while maintaining the computational performance of the digital counterpart.
Using a digital twin approach, we trained a four-node HORN in silico for sequential MNIST classification and transferred the trained parameters to an analog electronic implementation. 
A set of custom error metrics indicated that the analog system is able to successfully replicate the dynamics of the digital model in most test cases.
However, despite the overall well-matching dynamics, when using the readout layer of the digital model on the data generated by the analog system, we only observed $28.39\%$ agreement with the predictions of the digital model.
An analysis shows that this mismatch is due to a precision difference between the analog hardware and the floating-point representation exploited by the digital model to perform classification tasks.
When the analog system was utilized as a reservoir with a re-trained linear readout, its classification performance could be recovered to that of the digital twin, indicating preserved information content within the analog dynamics.
This proof-of-concept establishes that analog electronic circuits can effectively implement oscillatory neural networks for computation, providing a demonstration of energy-efficient analog systems that exploit brain-inspired transient dynamics for computation.
\end{abstract}


\maketitle

\section{Introduction}\label{intro}

Physical neural networks (PNNs) represent an emerging class of computational systems that leverage physical processes for information processing~\cite{wright_deep_2022,markovic2020physics,Yu2024}.
PNNs are also closely related to neuromorphic and reservoir computing~\cite{Schuman2022, Dalgaty2024}

In PNNs, non-digital physical systems are engineered such that their natural dynamics serve as the primary computational mechanism to perform machine learning tasks such as classification and regression~\cite{Aifer2025SolvingCompute}.
Various physical platforms have been investigated for the implementation of PNNs. These include analog electronic circuits~\cite{Merolla2014,zhong_memristor-based_2022,Dalgaty2024,Vandoorne2014}, photonic systems~\cite{Mirasso_photonics_review,wetzstein_inference_2020}, spintronic devices~\cite{romera_vowel_2018}, hybrid analog-digital architectures~\cite{chen_all-analog_2023}, and even purely mechanical devices~\cite{lee_mechanical_2022}.
These systems are designed such that their time-evolving states encode and transform information without relying on digital logic.
Two main methods are commonly used to train PNNs on a given task.
The first is the ``digital twin'' approach, in which the physical system is simulated digitally. The resulting digital model is then trained using a gradient based learning algorithm (backpropagation) and the learned parameters are transferred to the analog hardware~\cite{wright_deep_2022}.
The second method employs in situ techniques that bypass backpropagation, allowing for direct training on the physical substrate~\cite{science_backprop-free_prnn,Stern_2021}. A comprehensive discussion on the training of PNNs can be found in~\cite{momeni2024}.

In parallel to these advances in hardware, recurrent neural networks (RNNs) composed of coupled oscillators have been shown to possess compelling computational advantages over non-oscillating architectures~\cite{rusch_coupled_2021, Effenberger_doi:10.1073/pnas.2412830122}.
These oscillator-based architectures exhibit enhanced performance and parameter efficiency relative to conventional non-oscillating RNNs.
Specifically, the Harmonic Oscillator Recurrent Network (HORN) model was shown to reproduce many aspects of cortical dynamics while outperforming non-oscillating RNNs in parameter efficiency, learning speed, and robustness to perturbations~\cite{Effenberger_doi:10.1073/pnas.2412830122}.
These performance gains can be attributed to the unique dynamical features of oscillatory systems~\cite{Effenberger_doi:10.1073/pnas.2412830122}.
For example, phase-based encoding and synchronization enable robust information representation~\cite{fries2015,rosenblum_phase_2001}, while resonance and wave interference facilitate selective frequency filtering~\cite{hughes2019,buzsaki_mechanisms_2012}.
Moreover, the properties of fading memory in these systems facilitate temporal integration across multiple timescales~\cite{goltsev_critical_2013,DUBININ2024106179,Effenberger_doi:10.1073/pnas.2412830122}.
Unlike steady-state, typically attractor-based computational models~\cite{ashwin_excitable_2021}, HORNs utilize transient oscillatory dynamics for computation, mirroring the operational principles observed in biological neural systems~\cite{Nikolic_distributed_2009, Doelling_cortical_2015,SCHMIDT2023102796,Palmigiano2017}.
HORNs are biologically inspired by the ubiquity of oscillatory activity in the brain~\cite{fries2015,helfrich2018} and are particularly well-suited for realization in physical analog hardware~\cite{Csaba2020} where such dynamics can be implemented efficiently.

Although the computational benefits of oscillatory architectures have been demonstrated in digital simulations, their implementation in physical analog substrates remains limited.
To address this gap, we present an analog-electronic implementation of a HORN model on a hybrid analog-digital computer (anabrid Model 1~\cite{Model-1}) that can be programmed digitally, but performs all computations in an analog manner.
Implementing this solution involves addressing key challenges, such as precision limitations, dynamic range constraints, and parameter transfer protocols.
The proposed implementation employs a ``digital-twin'' approach~\cite{Chen2025}.
We first trained a HORN in silico on a sequential MNIST~\cite{lecun_mnist_2010} pattern recognition task, identifying optimal model parameters through supervised learning.
Subsequently, we mapped the optimal parameters determined by training to the analog hardware and inferred the $10,000$ MNIST samples from the test dataset using the analog implementation (see Fig. ~\ref{fig:network_sample} for an example).
To evaluate the fidelity of the analog implementation in replicating the dynamics of the digital model, we defined and computed comprehensive error metrics.
To assess the predictive accuracy of the analog implementation, we used two different readout mechanisms. 
We utilized the trained readout from the digital model within the analog implementation, but also investigated its potential as a reservoir~\cite{TANAKA2019100,Effenberger_doi:10.1073/pnas.2412830122,du2017}.

The remainder of this paper is organized as follows.
In Section~\ref{sec:model}, the Harmonic Oscillator Recurrent Network (HORN) model is introduced, including the adaptations necessary for its implementation in analog hardware.
Section~\ref{sec:Analog_implementation} outlines the experimental setup and the parameter-scaling procedure essential to transition the parameters from the digital model to the analog implementation.
In Section~\ref{sec:Results}, we assess the ability of the analog implementation to replicate the dynamics and task performance of its digital counterpart, while also examining its suitability as a reservoir.

\begin{figure*}
    \centering
    \includegraphics[width=\textwidth]{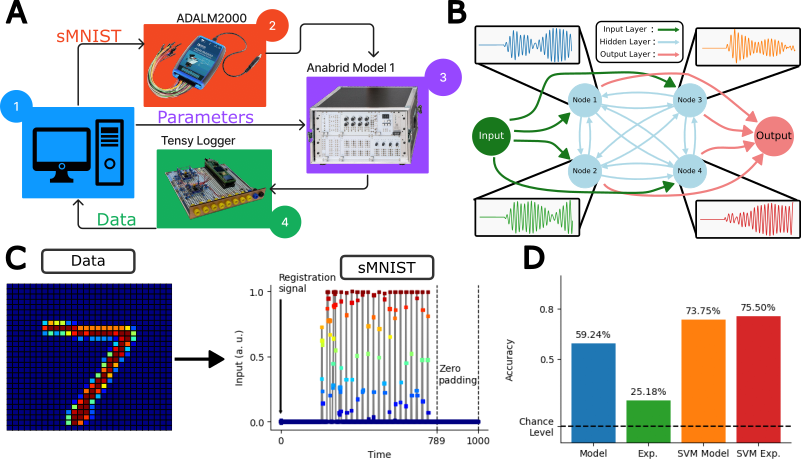}
    \caption{Experimental setup and performance comparison of digital and analog
    HORN implementations.
    \textbf{A.} Experimental setup with four components: digital computer (1), signal generator (2), analog computer (3), and data logger (4).
    \textbf{B.} Schematic of the four-node HORN model with example dynamics for each node (inset plots).
    \textbf{C.} MNIST preprocessing: original 28x28 sample (left) and sMNIST version with registration signal and zero-padding (right), time is in pixels, amplitudes are in a normalized grayscale (range [0,1]).
    \textbf{D.} Classification accuracy of digital model (blue), analog with digital readout (green), digital with SVM readout (orange), and analog reservoir with SVM readout (red).}
    \label{fig:1}
\end{figure*}

\section{Network Model}\label{sec:model} 
The Harmonic Oscillator Recurrent Network (HORN) model represents a recurrent neural network in which each node exhibits the dynamics of a damped harmonic oscillator (DHO unit)~\cite{Effenberger_doi:10.1073/pnas.2412830122}, see Fig.~\ref{fig:1}~B.
In a HORN consisting of $n$ nodes, the dynamics of each node $1\leq i\leq n$ is governed by the second-order ordinary differential equation (ODE)
\begin{equation}
    \ddot{x}_{i}(t)+ 2 \gamma_{i} \dot{x}_{i}(t) + \omega_{i}^2 x_{i}(t) = F(\mathbf{x},\dot{\mathbf{x}},t),
    \label{eq:DHO}
\end{equation}
where $t\in \mathbb{R}$ represents time, $x_{i}(t) \in \mathbb{R}$ denotes the state variable (that is, the amplitude) of the oscillator, $\omega_{i} > 0$ indicates the natural angular frequency, $\gamma_{i} > 0$ refers to the damping factor, and $F(\mathbf{x},\dot{\mathbf{x}},t)$ constitutes a forcing function defined subsequently. Furthermore, $\mathbf{x}(t) = (x_1(t), \dots, x_n(t))^T$ denotes the state vector of all oscillators in the network.
Note that the hyperparameters $\omega_i$ and $\gamma_i$ can vary independently. 
Together, they determine the relaxation dynamics of each node~\cite{kneubuhl_oscillations_1997, Effenberger_doi:10.1073/pnas.2412830122}.

The nodes are subjected to a time-varying forcing function
\begin{equation}
     F(\mathbf{x},\mathbf{\dot{x}},t) = \alpha \tanh \left( V\mathbf{x} + W\mathbf{\dot{x}} + \mathbf{I}s\left(t\right)\right),
\label{eq:forcing}
\end{equation}
where $\mathbf{I} \in \mathbb{R}^{1\times n}$ denotes the input weight matrix, that is, $I_{i}$ is a linear projection of the external input signal $s(t)\in \mathbb{R}$ to the $i$-th node.
The matrices $\mathbf{V}, \mathbf{W} \in \mathbb{R}^{n\times n}$ denote recurrent weight matrices.
Specifically, the entries $V_{ji}$ ($W_{ji}$) represent the strength of the amplitude (velocity) coupling from node $j$ to node $i$, see also~\cite{rusch_coupled_2021, Effenberger_doi:10.1073/pnas.2412830122}.
The diagonal entries of $\mathbf{V}$ ($\mathbf{W}$) represent the amplitude (velocity) feedback connection strengths of each node~\cite{Effenberger_doi:10.1073/pnas.2412830122}.

To accommodate the hardware limitations of the analog computer utilized in this study (see Sect.~\ref{sec:Analog_implementation}), we simplify the model to allow its implementation.
First, to accommodate the limited number of computational elements (adders, integrators, and coefficients), we chose a model consisting of $n=4$ nodes.
For the same reason, we also eliminate all feedback connections ($\mathbf{V}_{ii} = 0$, $\mathbf{W}_{ii} =0$).
Second, since an implementation of the $\operatorname{tanh}$-nonlinearity is not available, we excluded this function from the forcing function (Eq.~\ref{eq:forcing}).
Third, all recurrent amplitude coupling magnitudes ($\mathbf{V}$) are constrained to the interval $[0,1]$, prohibiting both inhibitory (negative) and amplifying couplings.
The restriction to non-negative values is dictated by circuit limitations, while the prohibition of amplifying couplings aims to prevent runaway dynamics.
These constraints enable the implementation of the network on the Model-1 analog computer utilized in this study~\cite{Model-1}.
For completeness, we also conducted tests on a velocity-coupled network in which we removed all amplitude couplings ($\mathbf{V}=0$) and velocity feedback connections ($W_{ii} = 0$).
Details are provided in Appendix~\ref{apx:velocity-network}.
We found that the amplitude and velocity coupled networks behaved similarly and only report on the findings for the amplitude coupled network in the main text.

We employ a digital twin approach to obtain network parameters for a sequential MNIST digit classification task by simulating the network on a digital computer using a time-discretized version of the HORN model.
To simulate the HORN model on a digital computer, we begin by converting the second-order ordinary differential equation (ODE) (Eq.~\ref{eq:DHO}) into a family of first-order ODEs through the introduction of an auxiliary variable $\dot{x} = y$. Subsequently, we develop a time-discrete representation of the resulting system using a microscopic time constant $h$, as outlined in~\cite{Effenberger_doi:10.1073/pnas.2412830122}.
Integrating this system through an Euler integration scheme and imposing the experimental constraints leads to the network update equations
\begin{equation}\label{eq:update_equations}
\begin{split}
    x_{i, t+1} =& x_{i, t} + h y_{i, t+1},\\
    y_{i, t+1} =& y_{i, t} + h\left[  \sum^{n}_{j \neq i}{\left(V_{ji}x_{i, t}\right)} + I_{i}s(t) - 2\gamma y_{i, t}  - \omega_{i}^{2} x_{i, t} \right],
\end{split}
\end{equation}
where $i$ denotes the node index and $t$ is a discrete-time index, see~\cite{Effenberger_doi:10.1073/pnas.2412830122}.
Note that a symplectic Euler integration~\cite{hairer_geometric_2003} is needed here to guarantee numerical stability, since the system is stiff~\cite{legenstein_spatiotemporalprocessingspiking}.

When simulating the discrete time model (Eq.~\ref{eq:update_equations}), we establish a connection between the time scales of the input and the system by presenting one pixel (intensity value) of an sMNIST stimulus $s(t)$ per iteration step, effectively setting $h=1$ in Eq.~\ref{eq:update_equations}.
The choice of $h=1$ allows us to express all time-related quantities in ``pixel'' time units, improving the interpretability of model quantities and hyperparameters.
In particular, this holds for the natural frequency parameters $\omega_i$ that can then be specified in radians per pixel, and the associated period ${\tau= 2\pi/(\omega h)}$.
Note that the system remains invariant under changes in $h \to h'$, provided that the system parameters are appropriately rescaled according to
\begin{equation}
\label{eq:rescaling}
\gamma'=c\gamma,\quad \omega'=c\omega,\quad \alpha'=c^{2}\alpha,
\end{equation}
where $c=h/h'$, see~\cite{Effenberger_doi:10.1073/pnas.2412830122}.
The frequency ${f_{\text{N}}={1}/{2h}}$ represents the upper limit of observable frequencies within the system, the \textit{Nyquist} frequency.
In practice, networks should not operate at frequencies near $f_{\text{N}}$, as the numerical integration error increases when approaching this limit.

The networks were trained on a sequential MNIST (sMNIST) handwritten digit classification task, a common benchmark for RNNs~\cite{rusch_coupled_2021, Effenberger_doi:10.1073/pnas.2412830122}.
For sMNIST, the 28x28 pixels MNIST samples are serialized into a time series of length $28\cdot 28 = 784$ pixels by collecting the intensity values from top left to bottom right in scanline order (Fig.~\ref{fig:1}~C).
An affine readout is performed at the last time step of stimulus presentation ($t = 784$ pixel) to perform digit classification, with $\mathbf{M} \in \mathbb{R}^{n\times n}$ denoting the readout matrix, and $b \in \mathbb{R}^{n} $ as the output bias vector.
The model was implemented in PyTorch~\cite{paszke_pytorch_2019} and the backpropagation through time (BPTT) algorithm was used to train all model parameters ($\mathbf{I}$, $\mathbf{V}$, $b$, and $\mathbf{M}$).
Following a digital twin approach~\cite{wright_deep_2022}, the learned weights (except $\mathbf{M}$ and $b$) are subsequently transferred to the analog implementation of the model.

\section{Experimental Setup}\label{sec:Analog_implementation}

This section outlines the hardware utilized (Sect.~\ref{subsec:hardware}), the circuit architecture for each DHO unit in the analog implementation (Sect.~\ref{subsec:circuit}), and the methodology for data preparation and presentation to the analog framework (Sect.~\ref{subsec:time_scale}).
Additionally, we also show how the process by which model parameters are scaled between the digital simulation and the analog implementation (Sect.~\ref{subsec:scaling}).

\subsection{Hardware Setup}\label{subsec:hardware} 

The experimental setup (Fig.~\ref{fig:1}~A) consists of four main parts:  (i) a digital computer, (ii) a signal generator, (iii) an analog circuit, and (iv) a data logger.

The digital computer (i) (Fig.~\ref{fig:1}~A, item 1) controls the experiment and performs three different tasks as described below.
First, it normalizes the model parameters and configures the analog circuit (iii) (see Sect.~\ref{subsec:scaling}).
Second, it normalizes the sMNIST input samples and transmits them to the signal generator (ii) (see Sect.~\ref{subsec:time_scale}).
Third, it collects the recorded data from the data logger (iv) following each inference run for subsequent analysis.

The signal generator (ii) is an ADALM2000~\cite{ADALM2000} (Fig.~\ref{fig:1}~A, item 2).
This device receives standardized sMNIST data (Fig.~\ref{fig:1}~C) from the digital computer (i) through a USB connection. Utilizing its signal generator functionality, it converts the digital data into an analog voltage trace, which is then fed to the analog circuit (iii).

The analog circuit (iii) was implemented using an anabrid Model-1 hybrid analog-digital computer~\cite{Model-1}(Fig.~\ref{fig:1}~A, item 3).
Each network node was implemented as a damped harmonic oscillator (Fig.~\ref{fig:oscillator_circuit}, see Sect.~\ref{subsec:circuit}).

The Teensy Logger~\cite{anabrid2025teensylogger} (iv) (Fig.~\ref{fig:1}~A) is custom hardware interfacing with the digital computer (i), the signal generator (ii), and the analog circuit (iii). It connects to the digital computer via USB and provides 5 analog input channels that are used to record the voltage traces of all four nodes as well as the input signal (to allow for precise temporal registration of the input and the nodal dynamics). Data is stored internally until the digital computer retrieves it after each inference run.

\subsection{Analog Circuit} \label{subsec:circuit}

The HORN model was implemented on an anabrid Model-1 analog computer, a modern, modular analog computer~\cite{Model-1}.
An analog computer is composed of various computing elements, such as integrators, summers, multipliers, and coefficients, that can be interconnected to model and solve a given problem~\cite{Ulmann_analog_computing_2022, Ulmann+2020}.
Analog computers operate non-algorithmically, lacking explicit memory, instructions, and a central processing unit and are ideally suited to solve problems that can be described as systems of coupled differential equations~\cite{Ulmann_analog_computing_2022}.
The ``in memory computation'' implemented in analog computers is well suited to simulate recurrent dynamics and can overcome the conceptual slowness of simulating dynamics on digital von-Neumann systems for which time has to be discretized and each dynamics step sequentially requires loading, updating, and writing of system state.
This makes such a system an ideal substrate to model a HORN consisting of a recurrently coupled network of dampled harmonic oscillator units (Eq.~\ref{eq:DHO}).
Quantities in analog computers are typically represented in abstract machine units confined to the interval $[-1,1]$, see~\cite{Ulmann_analog_computing_2022}.

\begin{figure}
    \includegraphics[width=8cm]{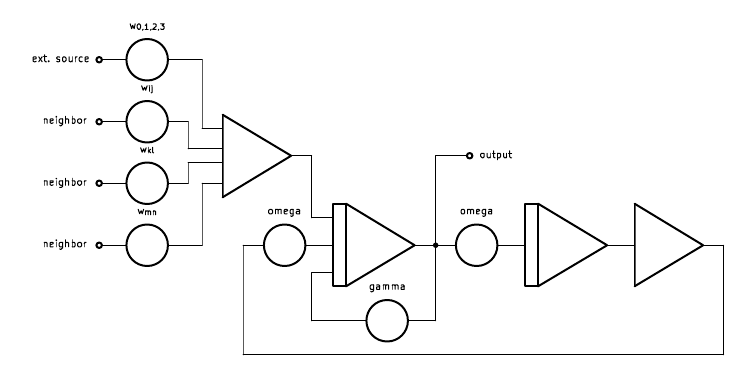}
    \caption{Analog circuit implementation of a damped harmonic oscillator node.
    The circuit contains integrators (triangular shapes with rectangles), summers (triangular shapes), and coefficients (circular elements) that can be digitally programmed to values in $[0,1]$.
    %
    % The core oscillator loop (lower right) consists of two integrators and
    % coefficients with value $\omega$, implementing the basic equation
    % $\ddot{x}=-\omega^2x$.
    % %
    % The leftmost integrator includes damping feedback through coefficient
    % $\gamma$.
    % %
    % The summer (upper left) combines the external input signal and coupling from
    % three neighboring oscillators.
    %
    Note that summers and integrators perform implicit sign inversion in this analog implementation.}
    \label{fig:oscillator_circuit}
\end{figure}

The circuit diagram of one of the four oscillators implemented in this study is presented in Fig.~\ref{fig:oscillator_circuit}.
The core oscillator circuit (lower right) consists of two integrators, a summer used for sign inversion, and two coefficients with value $\omega$.
This implements the basic differential equation of a harmonic oscillator, $\ddot{x}=-\omega^2x$.
The leftmost integrator has an additional feedback path from its output to one of its inputs by means of an additional coefficient of value $\gamma$.
Since the integrator performs an implicit change of sign, this feedback represents a damping term $\gamma$ controlling the decay rate of the oscillator.

Each oscillator has one output and four inputs, which are summed together using the summer shown in the upper left.
This summer serves a dual purpose: First, there are not enough inputs on the integrator to feed all inputs directly to it.
Second, the sign of the input signals must be flipped to have the oscillators run in phase when coupled.
The top input is the global excitation signal (``\textit{ext. source}'').
The three lower inputs are connected to neighboring oscillators.

The complete circuit implements an all-to-all connected HORN on four nodes that are coupled on their amplitudes using twelve coefficients, with an additional four coefficients for controlling the coupling to the external input signal.
See Appendix \ref{apx:velocity-network} for the velocity-coupled case.

\subsection{Physical Units}

Moving from the digital model to an implementation in analog hardware requires proper conversion between the units of the two systems and between discrete and continuous-time formulations.
This scaling is essential to ensure that the analog implementation replicates the dynamics observed in the digital simulation~\cite{Ulmann_analog_computing_2022}.
In analog hardware, the unit system is established by the characteristics of the component elements within the analog electronic circuit, such as the properties of integrators. In contrast, the digital model employs a more flexible unit system, as all quantities are represented by floating-point numbers that can be arbitrarily scaled according to Eq.~\ref{eq:rescaling}.
Since the digital model operates in discrete-time and the analog implementation operates in continuous-time, we first need to establish a relationship between the discrete and continuous-time parametrizations of the system.
Subsequently, we define the scaling relationships between all other model parameters.

\subsection{Temporal scale}\label{subsec:time_scale}

In contrast to digital simulations, where time is an abstract quantity, analog-electronic implementations necessitate running system dynamics for a defined duration $T$, measured in seconds.
Although the physical parameters of the analog circuit are fundamentally constrained only by the properties of its components (see Sect.~\ref{subsec:circuit}), accurately mapping the digital model to the analog implementation requires the selection of a parameter (for example, $\gamma$, $\omega$ or $h$) to serve as the basis for this mapping.
The selection of this parameter is arbitrary; however, it has to be made with the objective of replicating the dynamics of the chosen digital model. 
Consequently, once a parameter is established, all remaining parameters are derived from this selection. 
For example, if $\omega$ is designated as a basis, then $\gamma$ will be determined by the definition of $\omega$.
To achieve this, we fix the duration $T$ for presenting an sMNIST digit to the system in a way that meets several constraints, as described below. 
This choice is equivalent to setting the value of $h$ in the digital model, as establishing the total time is effectively the same as setting the pixel input duration in real time.
The critical quantity in this context is the machine integration factor $k_0$ of the integrators~\cite{Ulmann_analog_computing_2022}.
We opted to set the time scale of the analog computer by setting $k_0=10$, such that a value of $\omega_{max}=1$ (the maximal possible natural frequency of a node, in machine units; see Sect.~\ref{subsec:circuit}) corresponds to a frequency of $16$~Hz.
Since HORNs exploit resonance for feature extraction, optimal natural frequencies for sMNIST processing are near $\omega^{*}=\frac{2\pi}{28}$ (in rad/pixel units), as previously demonstrated~\cite{Effenberger_doi:10.1073/pnas.2412830122}.
Considering an analog implementation where $\omega^{*}$ corresponds to a maximal value of the frequency coefficient ($\omega_{\text{max}}=1$), the duration of the experiment must allow at least 28 cycles of a node, which corresponds to a minimal duration of $T_{\text{min}} = \frac{28}{16} = 1.75$~s.
To avoid inaccuracies of the analog circuit occurring at extreme parameter values, we set the experiment length to $T=6$~s, and scale system parameters as described in Sect.~\ref{subsec:scaling}.

Our experimental setup incorporates components operating at three distinct sampling frequencies (Fig.~\ref{fig:1}~A).
As a result, each system component yields a different total number of data points given a duration $T$ of an experimental run.
For instance, the number of stimulus samples for the sMNIST digit is $L$, the number of samples from the signal generator is $N_{I}$, and the number of samples recorded by the data logger is $N_{O}$ (Fig. \ref{fig:1}~A).
To ensure that the different components of the experiment have a coherent representation of time, these quantities should be commensurate, preferably they should satisfy $\frac{N_{I}}{L}, \frac{N_{O}}{L} \in \mathbb{N}$.
For example, given $L$ and $N_{O}$, let $\frac{N_O}{L} = a$.
Thus, $a$ data points recorded by the data logger correspond to the time period associated with one pixel of the digital model.

To allow for a registration of the signal generator and the data logger in the recorded data, a registration signal is added at the beginning of each sMNIST sample (Fig.~\ref{fig:1}~C), as the data logger records the input generated by the signal generator (Fig.~\ref{fig:1}~A).
Moreover, we zero-pad the sMNIST samples at the end to obtain samples of length $L=1000$ pixels, as this simplifies the calculations of all time-related quantities (Fig.~\ref{fig:1}~C).

The signal generator internally operates at a frequency of ${f_{I} = 75}$~MHz~\cite{ADALM2000}.
To present the standardized input consisting of ${L = 1000}$ data points (Fig.~\ref{fig:1}~D) within the time window $T$, the signal generator repeats each input data point ${S_{r} = 450 000}$ times, resulting in a total of $N_{I} = 450 \times 10^{6}$ input data points.

The data logger has the capacity to store up to $16,535$ samples across five recording channels~\cite{anabrid2025teensylogger}.
Since we utilize all five recording channels (input and amplitudes of four oscillators), recording a single time step requires storing 5 samples (we call this a \emph{data point}).
This limits the maximum number of data points to $16,535 / 5 = 3,307$.
To ensure that we can record an entire run, we configure the data logger with a sampling interval of $\tau_{O} = 3$~ms, resulting in $N_{O} = 2000$ data points per run.
This choice also ensures that the temporal sampling is not too coarse, resulting in two samples within the duration of each input sMNIST pixel.
The parameters are summarized in Table~\ref{tab:General_parameters}.

\begin{table}
    \begin{tabular}{|c|c|c|}
        \hline
        Symbol & Description & value\\
        \hline
        $L$ & sMNIST sample size & $1000$ pixels\\
        \hline
        $T$ & Total time of the experiment & $6 000 $~ms\\
        \hline
        $f_{I}$ & Sampling frequency of the input & $75 000 000 $~Hz\\
        \hline
        $S_{r}$ & Sample repetition & 450 000 \\
        \hline
        $N_{I}$ & Number of input data points & $ 450 \times 10^{6}$\\
        \hline
        $\tau_{O}$ & Output Sample interval & $3$~ ms \\
        \hline
        $N_{O}$ & Number of output data points & $2 000$ \\
        \hline
        $k_{0}$ & Computational integration factor & 10 \\
        \hline
    \end{tabular}
    \caption{ Experimental parameters for the analog HORN implementation.
    % Summary of temporal, sampling, and hardware configuration parameters used
    % to synchronize the digital control system, signal generation, and data
    % acquisition components.
    }\label{tab:General_parameters}
\end{table}

\subsection{Scaling}\label{subsec:scaling}

This section describes the process of mapping all model quantities to machine units, which is essential for the operation of the analog computer.
Analog computers typically code variables in the range $[-1,1]$ and parameters in the range $[0,1]$, and physical dimensions are expressed in ``machine units'', see~\cite{Ulmann_analog_computing_2022}.
Values of variables that exceed these limits result in clipping, which should be avoided to prevent this will lead to invalid results.
In what follows, let $\Delta S = T/L$ denote the sample duration (Table~\ref{tab:General_parameters}), let $k_{0}$ denote the machine integration factor~\cite{Ulmann_analog_computing_2022} (see Sect.~\ref{subsec:circuit}), let $\omega$ denote the natural frequency, let $\gamma$ denote the damping factor, and let $V_{(i,j)}$ denote the coupling strength from node $i$ to $j$.
To differentiate between the parameters of the digital model (defined in discrete time) and the ones of the analog implementation (defined in continuous time), we use subscripts ``M'' and ``E'', respectively.
The scaling relationship between the parameters of the digital model and their corresponding experimental values is expressed by
\begin{equation}
\begin{split}
    c &= \Delta S k_{0} \\
     I_{E} &= \frac{1}{c}\frac{ I_{M}}{\omega_{M}} \\
     \gamma_{E} &= \frac{\gamma_{M}}{c} \\
     \omega_{E} &= \frac{\omega_{M}}{c} \\
     V_{(i,j),E} &= \frac{1}{c}\frac{V_{(i,j),M}}{\omega_{j,M}}.
\end{split}
\end{equation}

% Here, $\omega_{j,M}$ denotes the natural frequency of the source node $j$ in the
% digital model; this convention matches the physical mapping of the coupling
% coefficient from node $j$ to node $i$ in the analog circuit.
%

The final quantity that requires scaling is the input matrix $I$. This matrix governs the magnitude of the input drive applied to each node, as described in Equation \ref{eq:forcing}, and consequently determines the dynamic range of the network.
%
%Therefore, $I$ has a direct influence on the distribution of the amplitude values of the network nodes, that is, the dynamic range of the network.
%
To prevent clipping, it is necessary to scale the entries of $I$, ensuring that the dynamic range of all nodes in the network remains within the valid range $\pm 1$ (in machine units) for each experimental run.
Theoretically, a single scaling of the entries of $I$ would suffice, given the maximal dynamic range across all runs.
However, this approach may lead to very small dynamic ranges for some experimental runs.
Due to limitations in machine precision (in our setup, it is $\Delta = \pm 0.03$ ~\cite{Model-1}), the analog computer may not accurately reproduce the dynamics of experimental runs characterized by a limited dynamic range.
Consequently, set-and-forget scaling may compromise the model performance of the analog implementation.

To achieve optimal model performance in the analog implementation, we distinctly scale the entries of $I$ for each experimental run. 
This approach guarantees that the effective dynamic range of the analog computer is maximized given the sample-specific dynamic range of the digital model.

The scaling of $I$ is performed through a sequence of simulations utilizing the digital model. In this process, we modify the scaling of the input matrix using a scalar multiplier to achieve maximal nodal amplitudes in the analog implementation, while ensuring that these amplitudes remain within the valid range $[-1,1]$ (see Fig.~\ref{fig:rescale-scheme}, Appendix~\ref{apx:Scaling_algorithm}).
This scaling procedure exemplifies a hybrid computing approach~\cite{Ulmann+2020}.

In the rescaled system, the variables, specifically the amplitudes $x$ (Eq.~\ref{eq:update_equations}), will be within the dynamic range of the analog machine. This observation is illustrated in Fig.~\ref{fig:network_sample} and Fig.~\ref{fig:Network_samples}.
%

%%%%%%%%%%%%%%%%%%%%%%%%%%%%%%%%%%%%%%%%%%%%%%%%%%%%%%%%%%%%%%%%%%%%%%%%%%%%%%%%%%%%%%%%%%%%%%%%%%%%%%%%%%%%
\section{Results}\label{sec:Results}% - The Analog Twin}

\begin{figure*}
    \centering
    \includegraphics[width=\linewidth]{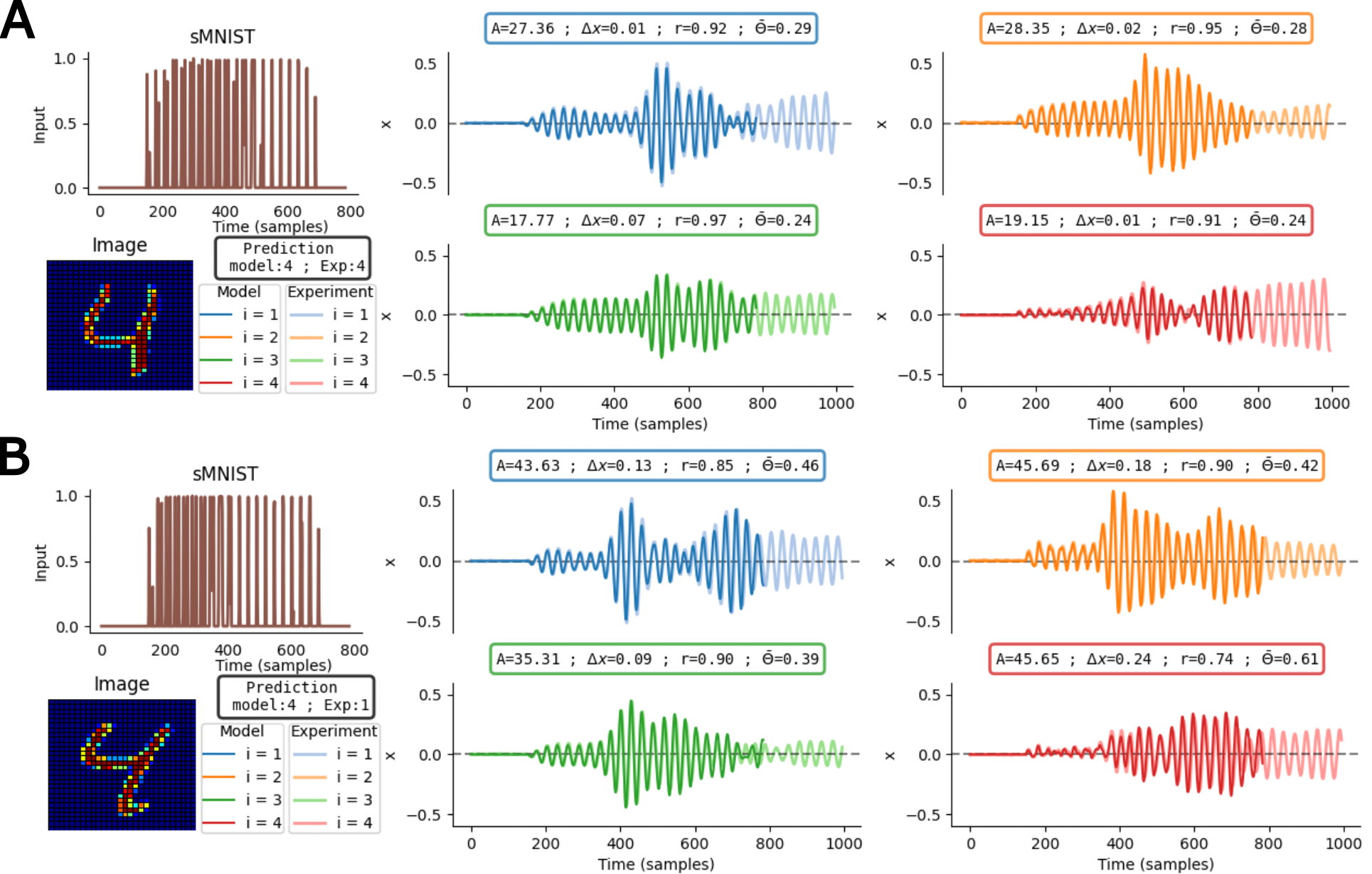}%{Images/BMT_3f79b084_sample_4.png}
    \caption{Comparison of analog and digital HORN dynamics for representative sMNIST samples demonstrating different reproduction outcomes.
    First column shows input sMNIST sample (top) and corresponding MNIST image (bottom).
    Subsequent columns display the dynamics of each of the four nodes, with light traces showing analog implementation dynamics and dark traces showing digital twin dynamics. 
    Time is measured in pixels for the digital twin and mapped to seconds in the analog implementation (T=6s). 
    Amplitudes are in machine units (with range $[-1,1]$).
    Error metrics quantify reproduction fidelity: $A$ (area between traces), $\Delta x$ (amplitude mismatch at $T=789$ pixel), $r$ (correlation), and $\overline{\Theta}$ (average phase difference).
    \textbf{A.} Case with matching predictions.
    \textbf{B.} Case with mismatched predictions.}
    \label{fig:network_sample}
\end{figure*}

Given the limited number of computational elements of the anabrid Model 1 analog computer used for this study, we evaluated the proposed setup utilizing a HORN model comprising $4$ nodes (see Fig.~\ref{fig:network_sample}).
We constructed a homogeneous HORN in which all nodes have the same values for the natural frequency $\omega_{M} = 0.22$ and the damping $\gamma_{M} = 0.01$.
The model was trained in silico using BPTT~\cite{Effenberger_doi:10.1073/pnas.2412830122}, achieving a classification accuracy of $59.24\%$ on sMNIST (Fig.\ref{fig:1}.~D).
%
% \rev{This performance significantly exceeds chance level (10\%) and
% demonstrates effective learning with minimal resources: only 4 nodes and 16
% trainable parameters transferred to analog hardware, representing a 350-fold
% reduction in parameters compared to typical MNIST classifiers \tocite.
% %
% For context, achieving $>50\%$ accuracy on sMNIST with such extreme parameter
% constraints demonstrates efficient information processing, particularly
% relevant for energy-constrained analog implementations where parameter
% efficiency is critical.}{Move to discussion?, find comparison for the HORN
% with other models}
%

%After determining optimal network parameters through in-silico training of the digital twin, we use these parameters as a basis for our analog implementation (the analog twin, see Sect.~\ref{subsec:scaling}).
Following training, we use these parameters in our analog implementation, referred to as the analog twin (see Section~\ref{subsec:scaling}).
%
% After setting up the analog implementation, 
We then performed inference on the $10,000$ sMNIST test samples using the analog twin (Fig.~\ref{fig:network_sample}).
To perform the inference, we processed the recorded time series data from all nodes through an affine readout layer. This layer maps the configuration of nodal amplitudes at pixel $T=789$ to predictions of MNIST digits (0-9) (see Fig.~\ref{fig:1}~C).
Applying the readout weights from the digital model to the data produced by the analog implementation resulted in a 28.39\% correspondence with the predictions of the digital model.
Among the $2,839$ samples for which both the digital and the analog twin predicted the same label, $1,973$ corresponded to the ground-truth MNIST label (Fig.~\ref{fig:apx_venn}).
We will show that this limited agreement between the predictions of the digital and the analog twin when using the readout layer of the digital model results from the digital model's reliance on high-precision floating-point representations.
Furthermore, we demonstrate that the analog twin preserves the dynamics of its digital counterpart. Using alternative readout strategies, we can implement a readout on the analog-generated data using a newly trained linear SVM, which recovers the performance level of the analog twin to the same extent as that of its digital counterpart.

\begin{figure*}
    \includegraphics[width=1\linewidth]{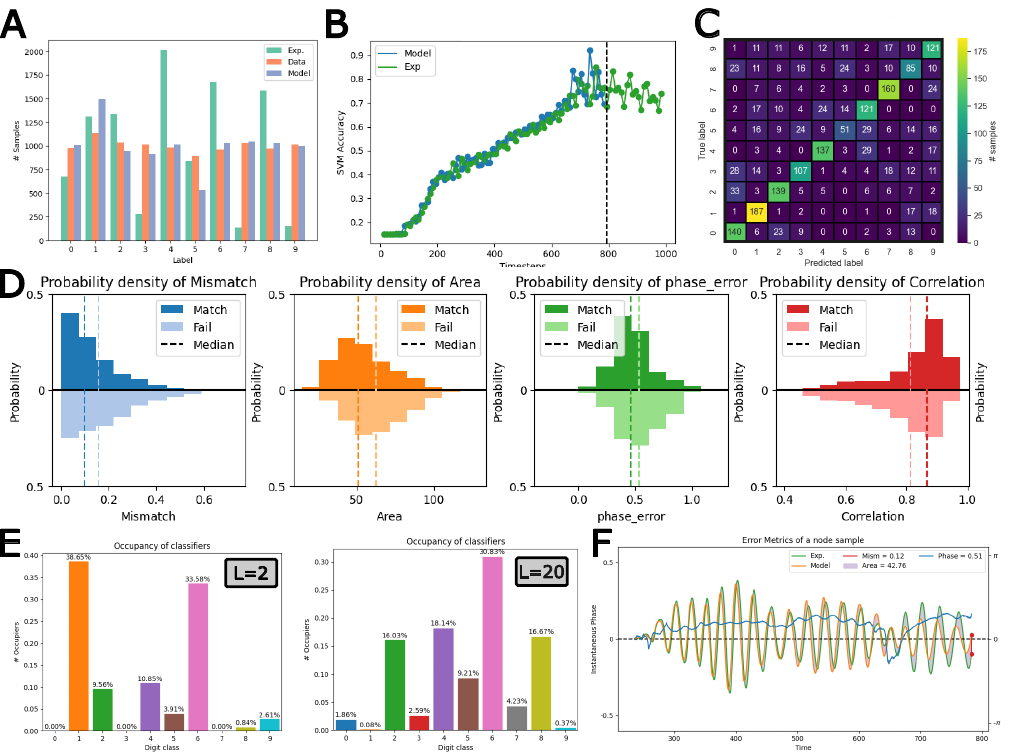}%{Images/Fig3_v3.png}
    \caption{
    % Comprehensive performance analysis revealing precision-limitedagreement
    % but preserved information content.
    Analysis of the analog implementation of the HORN model.
    \textbf{A.} Distribution of predicted labels showing class-specific differences: analog implementation (Exp.), digital twin (Model), and true MNIST labels (Data).
    Digits 0, 3, 7, and 9 are underrepresented in the analog implementation due to precision limitations.
    \textbf{B.} SVM readout accuracy over time demonstrating that both digital and analog systems achieve similar reservoir performance ($\approx74\%$) when appropriate readouts are used, confirming preserved information content.
    \textbf{C.} Confusion matrix for the analog implementation with a  SVM readout at $T = 789$ pixels, showing recovery of all digit classification capabilities when trained directly on analog dynamics.
    \textbf{D.} Error metric distributions comparing cases where analog and digital predictions match (top, low error) versus fail to match (bottom, higher error).
    Mismatch ($\Delta x$), area ($A$), phase ($\overline{\Theta}$), and correlation ($r$) metrics; dashed lines indicate medians.
    \textbf{E.} Volume occupancy analysis in 4D amplitude space $(x_1, x_2, x_3, x_4)$ illustrates precision limitations: small label volumes (digits 0, 3, 7, and 9) require higher precision than analog hardware provides.
    Left: $L=2$ (dense grid near origin); Right: $L=20$ (wide dynamic range).
    \textbf{F.} Error metrics visualization showing analog (green) vs digital (orange) node dynamics in the $[5,789]$ pixel time interval.
    Displayed error metrics are: area difference (purple), amplitude mismatch (red), and instantaneous phase difference (blue trace, right axis).}
    \label{fig:4}
\end{figure*}

%First, we examined how well the analog model reproduces the dynamics of the digital twin (Fig.~\ref{fig:network_sample}), to systematically evaluate this, we introduce the following error metrics, which are computed per node (see Fig.~\ref{fig:4}~F):
To systematically evaluate the degree to which the analog model replicates the dynamics of the digital twin (Fig.~\ref{fig:network_sample}), we introduce several error metrics calculated on a per-node basis (see Fig.~\ref{fig:4}~F):
(a) ``Mismatch'': difference in nodal amplitudes between digital and analog implementations at $T = 789$ pixel (see Fig.~\ref{fig:1}~C).
(b) ``Area'': difference in total area under time series of nodal amplitude.
(c) ``Phase'': average instantaneous phase difference between the time series.
(d) ``Correlation'': temporal correlation between the time series.
Here, (b), (c), and (d) are computed on the interval $[5,789]$ pixel (see Fig.~\ref{fig:1}~C).

After computing the error metrics for all nodes across $10,000$ test samples, we analyzed the distributions in two scenarios: where the label predictions of the analog and digital twins align, termed ``Match'', and where they do not, termed ``Fail'' (Fig.~\ref{fig:Error_samples}). Network-scale metrics were derived by averaging nodal metrics (Fig.~\ref{fig:4}~D).
%
%As expected, lower error metric values correlate with higher agreement between analog and digital predictions (Fig.~\ref{fig:4}~D), but, even when the digital twin's prediction is incorrect, we also observe dynamics of the analog implementation that fairly reproduces that of its digital twin (Fig.~\ref{fig:network_sample}).
%
As expected, lower values of the error metrics were associated with greater agreement between analog and digital predictions (Fig.~\ref{fig:4}~D). Notably, even when the digital twin produces an incorrect prediction, the dynamics of the analog implementation frequently replicate those of the digital twin (Fig.~\ref{fig:network_sample}).
Visualization of the error metrics (Fig.~\ref{fig:4}~F) reveals that inference runs with error values near the median of these metrics (Fig.~\ref{fig:4}~D) correspond to cases where the analog implementation successfully replicates the dynamics of its digital counterpart.
Overall, these fidings indicate that the analog twin is capable of reproducing the dynamics of the digital model in most cases.
%

% To evaluate prediction performance of the analog implementation, we processed
% the recorded time series data from all nodes using an affine readout layer
% that maps the configuration of nodal amplitudes at the final time step to
% MNIST digit predictions (0-9).
% %
% Using the affine readout parameters from the digital model on analog-generated
% data yielded matching predictions for 28.39\% of test samples.
% %
% This limited agreement results from the digital model's reliance on
% high-precision floating-point representations.
% %
% However, using alternative readout strategies, we demonstrate that the full
% performance of the digital model can be recovered in the analog
% implementation, as will be shown below.
% %
% Of the $2,839$ samples where the digital and analog twin predict the same
% label, $1,973$ were correct (i.e., the prediction matches the ground truth
% MNIST label, Fig.~\ref{fig:apx_venn}).
%

To better understand the differences between the predictions of the digital model and its analog twin, we analyzed the distribution of predicted labels for each digit class (Fig.~\ref{fig:4}A).
We find that at the digit class level, the analog implementation systematically fails to predict some labels (namely 9, 7, and 3), whereas these labels are predicted correctly by the digital model.
%
%Therefore, some digit classes can be correctly predicted by the digital model for most samples, but not by the analog twin.
%
This phenomenon warrants further analysis.
Comparing the predictions by the digital and the analog twin results in four possible scenarios: both twins accurately predicting the digit, both failing to predict correctly, or one twin making a correct prediction while the other fails (Fig.~\ref{fig:apx_venn}).
This study primarily investigates the capacity of the analog twin to replicate the dynamics of the digital twin. Consequently, we concentrate on instances where both twins produce identical predictions, irrespective of their correctness with respect to the ground truth.

To better understand the reasons of the limited agreement between the predictions of the digital and the analog twin (28.39\%, Fig.~\ref{fig:apx_venn}), we analyzed the volume that each label occupies in the four-dimensional amplitude space $(x_1,x_2,x_3,x_4)$ at the time step used for readout (referred to as decision space).
The volume associated with each label represents the subspace where the network predicts a specific digit class.
To facilitate this analysis, we created a four-dimensional grid by discretizing each dimension $x_1$, $x_2$, $x_3$, $x_4$ into $N=75$ evenly spaced bins, thereby covering the interval $[-L/2, L/2]$ in each dimension.
We considered two lattice sizes: $L=10$, which spans a cube corresponding to the maximum amplitude of any node in the analog system computed across all samples, and $L=2$, which produces a smaller lattice with a higher density of points near the origin $(0,0,0,0)$.

We observed that certain digit classes (0, 3, 7, and 9) occupy significantly smaller volumes within the decision space compared to other digits (Fig.~\ref{fig:4}~E).
Accessing these regions with small volumes necessitates high numerical precision. This indicates that the digitally trained network exploits floating-point precision capabilities for making its predictions. 
However, the degree of precision of the digital twin is orders of magnitude higher than the precision of the analog implementation. Therefore, some labels with small volumes in the decision space are not accessible to the analog twin (Fig.~\ref{fig:original_decoder_3f79b084}~B).
For this reason, we can conclude that the difference in precision between the digital and the analog twin is one of the factors that results in incorrect predictions, which explains the discrepancy in performance between the two twins.

However, this precision mismatch represents a solvable limitation rather than a fundamental constraint. Appropriate readout strategies can recover the performance of the analog implementation to the same degree as that of its digital counterpart, as demonstrated below.

To test whether the analog system preserves the same information that is present in the digital system in its dynamics, we discarded the affine readout layer trained as part of the digital model and replaced it with a linear Support Vector Machine (SVM) trained on the data generated by the analog implementation.
Thus, the analog twin was utilized as a reservoir~\cite{TANAKA2019100}.
For comparison and to avoid creating an unfair advantage for the analog twin, the same procedure was applied to its digital counterpart (Fig.~\ref{fig:4}~B).
The new setup with an SVM-based readout (Fig.~\ref{fig:1}~D) allows the analog twin to achieve the same level of performance as its digital counterpart (Fig.~\ref{fig:4}~B). In particular, it also recovers prediction accuracy across all digit classes (Fig.~\ref{fig:4}~C).
These results indicate that the analog twin preserves the information present in the digital model.
Furthermore, this observation indicates that analog neural networks can attain performance comparable to digital systems when utilized effectively.

\section{Discussion}\label{sec:Discussion}

This study showcases an implementation of the Harmonic Oscillator Recurrent Neural Network (HORN) model~\cite{Effenberger_doi:10.1073/pnas.2412830122} in analog electronic hardware, demonstrating that the analog implementation is able to reproduce the essential dynamics underlying model performance.
In particular, we employed a digital twin approach~\cite{wright_deep_2022} to train a small-scale HORN on a sequential MNIST (sMNIST) classification task. The learned parameters were subsequently transferred to an analog system for inference. 
For this proof of concept, we utilized a HORN model in its simplest configuration, characterized by a homogeneous network in which all nodes possess identical natural frequencies and damping coefficients. Additionally, we eliminated all feedback mechanisms present in the original HORN model.
Two network architectures were considered and both were found to faithfully reproduce the dynamics of its digital twin, an amplitude-couple one and a velocity-coupled one (see Appendix).

Quantitative analyses indicated that, despite hardware constraints such as a finite number of circuit components, a restricted dynamic range, and limited precision, the analog implementation was able to replicate the transient oscillatory dynamics of the digital model with high fidelity.
Despite the small size of the model and the additional simplifications imposed
by the limitations of the analog computer, the digital model achieved an
accuracy of $59.24\%$ on the sMNIST digit classification task using the low
number of 66 trainable parameters (of which only 16 were used in the analog
twin, and 50 belong to the readout layer).
This parameter efficiency of HORN models over other RNN architectures was
previously demonstrated~\cite{Effenberger_doi:10.1073/pnas.2412830122}, which
makes this model particularly well suited for implementation in
resource-constrained analog systems where parameter efficiency is crucial.

Error analysis indicates that the primary limitation of the analog model stems from the limited precision and dynamic range of its hardware, which results in reduced performance when applying the original digital readout weights to the analog model directly. 
This mismatch particularly affected MNIST digit classes whose
decision boundaries required fine-grained amplitude discrimination (namely, the digits 0, 3, 7, 9). 
Additional limitations stem from the simplified network used here, including the small
number of nodes, the absence of non-linear activation functions, and the restricted
connection weights, which were imposed by the limited number of circuit elements of the analog computer used in this study.

Importantly, these effects are not inherent to oscillatory analog computation, but reflect constraints of the readout strategy and hardware resolution.
When used as a reservoir with a separately trained readout provided by a linear SVM, 
the analog system achieved classification accuracy equivalent to the digital model 
(75.50\% vs 73.75\%, respectively), indicating that the dynamics of the analog system
preserves sufficient information for accurate decoding.
To address the limitations associated with readout, we propose that effective analog computation using oscillator networks may necessitate task-specific decoding strategies that exploit the full temporal structure of the network, rather than relying on fixed-time-point linear readouts~\cite{Ohkubo2024ReservoirComputing}.
Furthermore, we hypothesize that training with perturbed inputs or with intrinsic noise could enhance robustness and improve generalization.
However, it is important to note that the implementation of such strategies may require larger networks to maintain performance.

Our results have practical implications for the design of analog neural networks:
(i) We found that readout mechanisms may be more critical than internal dynamics for achieving target performance, suggesting that reservoir computing approaches may be particularly suited for analog implementations.
(ii) Our 16-parameter analog implementation demonstrates the feasibility of deploying neural networks in strongly resource-constrained environments where digital alternatives are impractical, such as edge computing applications~\cite{Satyanarayanan2017} or energy-limited sensors.
(iii) The implementation of the HORN model in this setting supports the broader proposition that oscillatory transient dynamics~\cite{Palmigiano2017} offers a viable substrate for robust, efficient, and real-time physical computation, in contrast to networks based on the principles of steady-state attractors~\cite{Amit_1989}.

The current work opens up several avenues for designing energy-efficient, brain-inspired computing architectures and hardware~\cite{indiveri2015}.
In particular, analog oscillator networks with parallelized information processing might benefit from neuromorphic computing approaches~\cite{Schuman2022}, potentially enabling fully analog learning systems~\cite{Kendall2020, Yu2024}.
Future work may explore larger networks~\cite{Kudithipudi2025}, the inclusion of nonlinearities, spiking networks that implement similar principles \cite{Baronig2024AdvancingSpatioTemporal, higuchi2024balanced}, and incorporating in situ learning methods to bypass the need for parameter transfer from a digital twin and thereby potentially enabling the construction of entirely analog systems capable of real-time, low-power machine learning.

In summary, the proof-of-concept presented here demonstrates that brain-inspired analog-electronic oscillator networks and their transient dynamics can serve as physically instantiated recurrent neural networks capable of performing machine learning tasks with extreme power efficiency~\cite{Aifer2025SolvingCompute}.
With continued advances in programmable analog hardware, oscillator-based recurrent networks could become a practical building block for real-time, energy-efficient, and fully analog neuromorphic computing systems.

\bibliography{sn-bibliography}%

%apsrev4-2.bst 2019-01-14 (MD) hand-edited version of apsrev4-1.bst
%Control: key (0)
%Control: author (8) initials jnrlst
%Control: editor formatted (1) identically to author
%Control: production of article title (0) allowed
%Control: page (0) single
%Control: year (1) truncated
%Control: production of eprint (0) enabled
\begin{thebibliography}{54}%
\makeatletter
\providecommand \@ifxundefined [1]{%
 \@ifx{#1\undefined}
}%
\providecommand \@ifnum [1]{%
 \ifnum #1\expandafter \@firstoftwo
 \else \expandafter \@secondoftwo
 \fi
}%
\providecommand \@ifx [1]{%
 \ifx #1\expandafter \@firstoftwo
 \else \expandafter \@secondoftwo
 \fi
}%
\providecommand \natexlab [1]{#1}%
\providecommand \enquote  [1]{``#1''}%
\providecommand \bibnamefont  [1]{#1}%
\providecommand \bibfnamefont [1]{#1}%
\providecommand \citenamefont [1]{#1}%
\providecommand \href@noop [0]{\@secondoftwo}%
\providecommand \href [0]{\begingroup \@sanitize@url \@href}%
\providecommand \@href[1]{\@@startlink{#1}\@@href}%
\providecommand \@@href[1]{\endgroup#1\@@endlink}%
\providecommand \@sanitize@url [0]{\catcode `\\12\catcode `\$12\catcode `\&12\catcode `\#12\catcode `\^12\catcode `\_12\catcode `\%12\relax}%
\providecommand \@@startlink[1]{}%
\providecommand \@@endlink[0]{}%
\providecommand \url  [0]{\begingroup\@sanitize@url \@url }%
\providecommand \@url [1]{\endgroup\@href {#1}{\urlprefix }}%
\providecommand \urlprefix  [0]{URL }%
\providecommand \Eprint [0]{\href }%
\providecommand \doibase [0]{https://doi.org/}%
\providecommand \selectlanguage [0]{\@gobble}%
\providecommand \bibinfo  [0]{\@secondoftwo}%
\providecommand \bibfield  [0]{\@secondoftwo}%
\providecommand \translation [1]{[#1]}%
\providecommand \BibitemOpen [0]{}%
\providecommand \bibitemStop [0]{}%
\providecommand \bibitemNoStop [0]{.\EOS\space}%
\providecommand \EOS [0]{\spacefactor3000\relax}%
\providecommand \BibitemShut  [1]{\csname bibitem#1\endcsname}%
\let\auto@bib@innerbib\@empty
%</preamble>
\bibitem [{\citenamefont {Wright}\ \emph {et~al.}(2022)\citenamefont {Wright}, \citenamefont {Onodera}, \citenamefont {Stein}, \citenamefont {Wang}, \citenamefont {Schachter}, \citenamefont {Hu},\ and\ \citenamefont {{McMahon}}}]{wright_deep_2022}%
  \BibitemOpen
  \bibfield  {author} {\bibinfo {author} {\bibfnamefont {L.~G.}\ \bibnamefont {Wright}}, \bibinfo {author} {\bibfnamefont {T.}~\bibnamefont {Onodera}}, \bibinfo {author} {\bibfnamefont {M.~M.}\ \bibnamefont {Stein}}, \bibinfo {author} {\bibfnamefont {T.}~\bibnamefont {Wang}}, \bibinfo {author} {\bibfnamefont {D.~T.}\ \bibnamefont {Schachter}}, \bibinfo {author} {\bibfnamefont {Z.}~\bibnamefont {Hu}},\ and\ \bibinfo {author} {\bibfnamefont {P.~L.}\ \bibnamefont {{McMahon}}},\ }\bibfield  {title} {\bibinfo {title} {Deep physical neural networks trained with backpropagation},\ }\href {https://doi.org/10.1038/s41586-021-04223-6} {\bibfield  {journal} {\bibinfo  {journal} {Nature}\ }\textbf {\bibinfo {volume} {601}},\ \bibinfo {pages} {549} (\bibinfo {year} {2022})}\BibitemShut {NoStop}%
\bibitem [{\citenamefont {Markovi{\'c}}\ \emph {et~al.}(2020)\citenamefont {Markovi{\'c}}, \citenamefont {Mizrahi}, \citenamefont {Querlioz},\ and\ \citenamefont {Grollier}}]{markovic2020physics}%
  \BibitemOpen
  \bibfield  {author} {\bibinfo {author} {\bibfnamefont {D.}~\bibnamefont {Markovi{\'c}}}, \bibinfo {author} {\bibfnamefont {A.}~\bibnamefont {Mizrahi}}, \bibinfo {author} {\bibfnamefont {D.}~\bibnamefont {Querlioz}},\ and\ \bibinfo {author} {\bibfnamefont {J.}~\bibnamefont {Grollier}},\ }\bibfield  {title} {\bibinfo {title} {Physics for neuromorphic computing},\ }\href@noop {} {\bibfield  {journal} {\bibinfo  {journal} {Nature Reviews Physics}\ }\textbf {\bibinfo {volume} {2}},\ \bibinfo {pages} {499} (\bibinfo {year} {2020})}\BibitemShut {NoStop}%
\bibitem [{\citenamefont {Yu}\ \emph {et~al.}(2024)\citenamefont {Yu}, \citenamefont {Guo}, \citenamefont {Xiao},\ and\ \citenamefont {Shen}}]{Yu2024}%
  \BibitemOpen
  \bibfield  {author} {\bibinfo {author} {\bibfnamefont {W.}~\bibnamefont {Yu}}, \bibinfo {author} {\bibfnamefont {H.}~\bibnamefont {Guo}}, \bibinfo {author} {\bibfnamefont {J.}~\bibnamefont {Xiao}},\ and\ \bibinfo {author} {\bibfnamefont {J.}~\bibnamefont {Shen}},\ }\bibfield  {title} {\bibinfo {title} {Physical neural networks with self-learning capabilities},\ }\href {https://doi.org/10.1007/s11433-024-2403-x} {\bibfield  {journal} {\bibinfo  {journal} {Science China Physics, Mechanics {\&} Astronomy}\ }\textbf {\bibinfo {volume} {67}},\ \bibinfo {pages} {287501} (\bibinfo {year} {2024})}\BibitemShut {NoStop}%
\bibitem [{\citenamefont {Schuman}\ \emph {et~al.}(2022)\citenamefont {Schuman}, \citenamefont {Kulkarni}, \citenamefont {Parsa}, \citenamefont {Mitchell}, \citenamefont {Date},\ and\ \citenamefont {Kay}}]{Schuman2022}%
  \BibitemOpen
  \bibfield  {author} {\bibinfo {author} {\bibfnamefont {C.~D.}\ \bibnamefont {Schuman}}, \bibinfo {author} {\bibfnamefont {S.~R.}\ \bibnamefont {Kulkarni}}, \bibinfo {author} {\bibfnamefont {M.}~\bibnamefont {Parsa}}, \bibinfo {author} {\bibfnamefont {J.~P.}\ \bibnamefont {Mitchell}}, \bibinfo {author} {\bibfnamefont {P.}~\bibnamefont {Date}},\ and\ \bibinfo {author} {\bibfnamefont {B.}~\bibnamefont {Kay}},\ }\bibfield  {title} {\bibinfo {title} {Opportunities for neuromorphic computing algorithms and applications},\ }\href {https://doi.org/10.1038/s43588-021-00184-y} {\bibfield  {journal} {\bibinfo  {journal} {Nature Computational Science}\ }\textbf {\bibinfo {volume} {2}},\ \bibinfo {pages} {10} (\bibinfo {year} {2022})}\BibitemShut {NoStop}%
\bibitem [{\citenamefont {Dalgaty}\ \emph {et~al.}(2024)\citenamefont {Dalgaty}, \citenamefont {Moro}, \citenamefont {Demira{\u{g}}}, \citenamefont {De~Pra}, \citenamefont {Indiveri}, \citenamefont {Vianello},\ and\ \citenamefont {Payvand}}]{Dalgaty2024}%
  \BibitemOpen
  \bibfield  {author} {\bibinfo {author} {\bibfnamefont {T.}~\bibnamefont {Dalgaty}}, \bibinfo {author} {\bibfnamefont {F.}~\bibnamefont {Moro}}, \bibinfo {author} {\bibfnamefont {Y.}~\bibnamefont {Demira{\u{g}}}}, \bibinfo {author} {\bibfnamefont {A.}~\bibnamefont {De~Pra}}, \bibinfo {author} {\bibfnamefont {G.}~\bibnamefont {Indiveri}}, \bibinfo {author} {\bibfnamefont {E.}~\bibnamefont {Vianello}},\ and\ \bibinfo {author} {\bibfnamefont {M.}~\bibnamefont {Payvand}},\ }\bibfield  {title} {\bibinfo {title} {Mosaic: in-memory computing and routing for small-world spike-based neuromorphic systems},\ }\href {https://doi.org/10.1038/s41467-023-44365-x} {\bibfield  {journal} {\bibinfo  {journal} {Nature Communications}\ }\textbf {\bibinfo {volume} {15}},\ \bibinfo {pages} {142} (\bibinfo {year} {2024})}\BibitemShut {NoStop}%
\bibitem [{\citenamefont {Aifer}\ \emph {et~al.}(2025)\citenamefont {Aifer}, \citenamefont {Belateche}, \citenamefont {Bramhavar}, \citenamefont {Camsari}, \citenamefont {Coles}, \citenamefont {Crooks}, \citenamefont {Durian}, \citenamefont {Liu}, \citenamefont {Marchenkova}, \citenamefont {Martinez}, \citenamefont {McMahon}, \citenamefont {Sbahi}, \citenamefont {Weiner},\ and\ \citenamefont {Wright}}]{Aifer2025SolvingCompute}%
  \BibitemOpen
  \bibfield  {author} {\bibinfo {author} {\bibfnamefont {M.}~\bibnamefont {Aifer}}, \bibinfo {author} {\bibfnamefont {Z.}~\bibnamefont {Belateche}}, \bibinfo {author} {\bibfnamefont {S.}~\bibnamefont {Bramhavar}}, \bibinfo {author} {\bibfnamefont {K.~Y.}\ \bibnamefont {Camsari}}, \bibinfo {author} {\bibfnamefont {P.~J.}\ \bibnamefont {Coles}}, \bibinfo {author} {\bibfnamefont {G.}~\bibnamefont {Crooks}}, \bibinfo {author} {\bibfnamefont {D.~J.}\ \bibnamefont {Durian}}, \bibinfo {author} {\bibfnamefont {A.~J.}\ \bibnamefont {Liu}}, \bibinfo {author} {\bibfnamefont {A.}~\bibnamefont {Marchenkova}}, \bibinfo {author} {\bibfnamefont {A.~J.}\ \bibnamefont {Martinez}}, \bibinfo {author} {\bibfnamefont {P.~L.}\ \bibnamefont {McMahon}}, \bibinfo {author} {\bibfnamefont {F.}~\bibnamefont {Sbahi}}, \bibinfo {author} {\bibfnamefont {B.}~\bibnamefont {Weiner}},\ and\ \bibinfo {author} {\bibfnamefont {L.~G.}\ \bibnamefont {Wright}},\ }\href {https://doi.org/10.48550/arXiv.2507.10463} {\bibinfo {title} {Solving the compute
  crisis with physics-based {{ASICs}}}} (\bibinfo {year} {2025}),\ \Eprint {https://arxiv.org/abs/2507.10463} {arXiv:2507.10463 [cs]} \BibitemShut {NoStop}%
\bibitem [{\citenamefont {Merolla}\ \emph {et~al.}(2014)\citenamefont {Merolla}, \citenamefont {Arthur}, \citenamefont {Alvarez-Icaza}, \citenamefont {Cassidy}, \citenamefont {Sawada}, \citenamefont {Akopyan}, \citenamefont {Jackson}, \citenamefont {Imam}, \citenamefont {Guo}, \citenamefont {Nakamura}, \citenamefont {Brezzo}, \citenamefont {Vo}, \citenamefont {Esser}, \citenamefont {Appuswamy}, \citenamefont {Taba}, \citenamefont {Amir}, \citenamefont {Flickner}, \citenamefont {Risk}, \citenamefont {Manohar},\ and\ \citenamefont {Modha}}]{Merolla2014}%
  \BibitemOpen
  \bibfield  {author} {\bibinfo {author} {\bibfnamefont {P.~A.}\ \bibnamefont {Merolla}}, \bibinfo {author} {\bibfnamefont {J.~V.}\ \bibnamefont {Arthur}}, \bibinfo {author} {\bibfnamefont {R.}~\bibnamefont {Alvarez-Icaza}}, \bibinfo {author} {\bibfnamefont {A.~S.}\ \bibnamefont {Cassidy}}, \bibinfo {author} {\bibfnamefont {J.}~\bibnamefont {Sawada}}, \bibinfo {author} {\bibfnamefont {F.}~\bibnamefont {Akopyan}}, \bibinfo {author} {\bibfnamefont {B.~L.}\ \bibnamefont {Jackson}}, \bibinfo {author} {\bibfnamefont {N.}~\bibnamefont {Imam}}, \bibinfo {author} {\bibfnamefont {C.}~\bibnamefont {Guo}}, \bibinfo {author} {\bibfnamefont {Y.}~\bibnamefont {Nakamura}}, \bibinfo {author} {\bibfnamefont {B.}~\bibnamefont {Brezzo}}, \bibinfo {author} {\bibfnamefont {I.}~\bibnamefont {Vo}}, \bibinfo {author} {\bibfnamefont {S.~K.}\ \bibnamefont {Esser}}, \bibinfo {author} {\bibfnamefont {R.}~\bibnamefont {Appuswamy}}, \bibinfo {author} {\bibfnamefont {B.}~\bibnamefont {Taba}}, \bibinfo {author} {\bibfnamefont
  {A.}~\bibnamefont {Amir}}, \bibinfo {author} {\bibfnamefont {M.~D.}\ \bibnamefont {Flickner}}, \bibinfo {author} {\bibfnamefont {W.~P.}\ \bibnamefont {Risk}}, \bibinfo {author} {\bibfnamefont {R.}~\bibnamefont {Manohar}},\ and\ \bibinfo {author} {\bibfnamefont {D.~S.}\ \bibnamefont {Modha}},\ }\bibfield  {title} {\bibinfo {title} {A million spiking-neuron integrated circuit with a scalable communication network and interface},\ }\href {https://doi.org/10.1126/science.1254642} {\bibfield  {journal} {\bibinfo  {journal} {Science}\ }\textbf {\bibinfo {volume} {345}},\ \bibinfo {pages} {668} (\bibinfo {year} {2014})}\BibitemShut {NoStop}%
\bibitem [{\citenamefont {Zhong}\ \emph {et~al.}(2022)\citenamefont {Zhong}, \citenamefont {Tang}, \citenamefont {Li}, \citenamefont {Liang}, \citenamefont {Liu}, \citenamefont {Li}, \citenamefont {Xi}, \citenamefont {Yao}, \citenamefont {Hao}, \citenamefont {Gao}, \citenamefont {Qian},\ and\ \citenamefont {Wu}}]{zhong_memristor-based_2022}%
  \BibitemOpen
  \bibfield  {author} {\bibinfo {author} {\bibfnamefont {Y.}~\bibnamefont {Zhong}}, \bibinfo {author} {\bibfnamefont {J.}~\bibnamefont {Tang}}, \bibinfo {author} {\bibfnamefont {X.}~\bibnamefont {Li}}, \bibinfo {author} {\bibfnamefont {X.}~\bibnamefont {Liang}}, \bibinfo {author} {\bibfnamefont {Z.}~\bibnamefont {Liu}}, \bibinfo {author} {\bibfnamefont {Y.}~\bibnamefont {Li}}, \bibinfo {author} {\bibfnamefont {Y.}~\bibnamefont {Xi}}, \bibinfo {author} {\bibfnamefont {P.}~\bibnamefont {Yao}}, \bibinfo {author} {\bibfnamefont {Z.}~\bibnamefont {Hao}}, \bibinfo {author} {\bibfnamefont {B.}~\bibnamefont {Gao}}, \bibinfo {author} {\bibfnamefont {H.}~\bibnamefont {Qian}},\ and\ \bibinfo {author} {\bibfnamefont {H.}~\bibnamefont {Wu}},\ }\bibfield  {title} {\bibinfo {title} {A memristor-based analogue reservoir computing system for real-time and power-efficient signal processing},\ }\href {https://doi.org/10.1038/s41928-022-00838-3} {\bibfield  {journal} {\bibinfo  {journal} {Nature Electronics}\ ,\ \bibinfo {pages}
  {672}} (\bibinfo {year} {2022})}\BibitemShut {NoStop}%
\bibitem [{\citenamefont {Vandoorne}\ \emph {et~al.}(2014)\citenamefont {Vandoorne}, \citenamefont {Mechet}, \citenamefont {Van~Vaerenbergh}, \citenamefont {Fiers}, \citenamefont {Morthier}, \citenamefont {Verstraeten}, \citenamefont {Schrauwen}, \citenamefont {Dambre},\ and\ \citenamefont {Bienstman}}]{Vandoorne2014}%
  \BibitemOpen
  \bibfield  {author} {\bibinfo {author} {\bibfnamefont {K.}~\bibnamefont {Vandoorne}}, \bibinfo {author} {\bibfnamefont {P.}~\bibnamefont {Mechet}}, \bibinfo {author} {\bibfnamefont {T.}~\bibnamefont {Van~Vaerenbergh}}, \bibinfo {author} {\bibfnamefont {M.}~\bibnamefont {Fiers}}, \bibinfo {author} {\bibfnamefont {G.}~\bibnamefont {Morthier}}, \bibinfo {author} {\bibfnamefont {D.}~\bibnamefont {Verstraeten}}, \bibinfo {author} {\bibfnamefont {B.}~\bibnamefont {Schrauwen}}, \bibinfo {author} {\bibfnamefont {J.}~\bibnamefont {Dambre}},\ and\ \bibinfo {author} {\bibfnamefont {P.}~\bibnamefont {Bienstman}},\ }\bibfield  {title} {\bibinfo {title} {Experimental demonstration of reservoir computing on a silicon photonics chip},\ }\href {https://doi.org/10.1038/ncomms4541} {\bibfield  {journal} {\bibinfo  {journal} {Nature Communications}\ }\textbf {\bibinfo {volume} {5}},\ \bibinfo {pages} {3541} (\bibinfo {year} {2014})}\BibitemShut {NoStop}%
\bibitem [{\citenamefont {Abreu}\ \emph {et~al.}(2024)\citenamefont {Abreu}, \citenamefont {Boikov}, \citenamefont {Goldmann}, \citenamefont {Jonuzi}, \citenamefont {Lupo}, \citenamefont {Masaad}, \citenamefont {Nguyen}, \citenamefont {Picco}, \citenamefont {Pourcel}, \citenamefont {Skalli}, \citenamefont {Talandier}, \citenamefont {Vettelschoss}, \citenamefont {Vlieg}, \citenamefont {Argyris}, \citenamefont {Bienstman}, \citenamefont {Brunner}, \citenamefont {Dambre}, \citenamefont {Daudet}, \citenamefont {Domenech}, \citenamefont {Fischer}, \citenamefont {Horst}, \citenamefont {Massar}, \citenamefont {Mirasso}, \citenamefont {Offrein}, \citenamefont {Rossi}, \citenamefont {Soriano}, \citenamefont {Sygletos},\ and\ \citenamefont {Turitsyn}}]{Mirasso_photonics_review}%
  \BibitemOpen
  \bibfield  {author} {\bibinfo {author} {\bibfnamefont {S.}~\bibnamefont {Abreu}}, \bibinfo {author} {\bibfnamefont {I.}~\bibnamefont {Boikov}}, \bibinfo {author} {\bibfnamefont {M.}~\bibnamefont {Goldmann}}, \bibinfo {author} {\bibfnamefont {T.}~\bibnamefont {Jonuzi}}, \bibinfo {author} {\bibfnamefont {A.}~\bibnamefont {Lupo}}, \bibinfo {author} {\bibfnamefont {S.}~\bibnamefont {Masaad}}, \bibinfo {author} {\bibfnamefont {L.}~\bibnamefont {Nguyen}}, \bibinfo {author} {\bibfnamefont {E.}~\bibnamefont {Picco}}, \bibinfo {author} {\bibfnamefont {G.}~\bibnamefont {Pourcel}}, \bibinfo {author} {\bibfnamefont {A.}~\bibnamefont {Skalli}}, \bibinfo {author} {\bibfnamefont {L.}~\bibnamefont {Talandier}}, \bibinfo {author} {\bibfnamefont {B.}~\bibnamefont {Vettelschoss}}, \bibinfo {author} {\bibfnamefont {E.}~\bibnamefont {Vlieg}}, \bibinfo {author} {\bibfnamefont {A.}~\bibnamefont {Argyris}}, \bibinfo {author} {\bibfnamefont {P.}~\bibnamefont {Bienstman}}, \bibinfo {author} {\bibfnamefont {D.}~\bibnamefont
  {Brunner}}, \bibinfo {author} {\bibfnamefont {J.}~\bibnamefont {Dambre}}, \bibinfo {author} {\bibfnamefont {L.}~\bibnamefont {Daudet}}, \bibinfo {author} {\bibfnamefont {J.}~\bibnamefont {Domenech}}, \bibinfo {author} {\bibfnamefont {I.}~\bibnamefont {Fischer}}, \bibinfo {author} {\bibfnamefont {F.}~\bibnamefont {Horst}}, \bibinfo {author} {\bibfnamefont {S.}~\bibnamefont {Massar}}, \bibinfo {author} {\bibfnamefont {C.}~\bibnamefont {Mirasso}}, \bibinfo {author} {\bibfnamefont {B.}~\bibnamefont {Offrein}}, \bibinfo {author} {\bibfnamefont {A.}~\bibnamefont {Rossi}}, \bibinfo {author} {\bibfnamefont {M.}~\bibnamefont {Soriano}}, \bibinfo {author} {\bibfnamefont {S.}~\bibnamefont {Sygletos}},\ and\ \bibinfo {author} {\bibfnamefont {S.}~\bibnamefont {Turitsyn}},\ }\bibfield  {title} {\bibinfo {title} {A photonics perspective on computing with physical substrates},\ }\href {https://doi.org/https://doi.org/10.1016/j.revip.2024.100093} {\bibfield  {journal} {\bibinfo  {journal} {Reviews in Physics}\ }\textbf
  {\bibinfo {volume} {12}},\ \bibinfo {pages} {100093} (\bibinfo {year} {2024})}\BibitemShut {NoStop}%
\bibitem [{\citenamefont {Wetzstein}\ \emph {et~al.}(2020)\citenamefont {Wetzstein}, \citenamefont {Ozcan}, \citenamefont {Gigan}, \citenamefont {Fan}, \citenamefont {Englund}, \citenamefont {Soljačić}, \citenamefont {Denz}, \citenamefont {Miller},\ and\ \citenamefont {Psaltis}}]{wetzstein_inference_2020}%
  \BibitemOpen
  \bibfield  {author} {\bibinfo {author} {\bibfnamefont {G.}~\bibnamefont {Wetzstein}}, \bibinfo {author} {\bibfnamefont {A.}~\bibnamefont {Ozcan}}, \bibinfo {author} {\bibfnamefont {S.}~\bibnamefont {Gigan}}, \bibinfo {author} {\bibfnamefont {S.}~\bibnamefont {Fan}}, \bibinfo {author} {\bibfnamefont {D.}~\bibnamefont {Englund}}, \bibinfo {author} {\bibfnamefont {M.}~\bibnamefont {Soljačić}}, \bibinfo {author} {\bibfnamefont {C.}~\bibnamefont {Denz}}, \bibinfo {author} {\bibfnamefont {D.~A.~B.}\ \bibnamefont {Miller}},\ and\ \bibinfo {author} {\bibfnamefont {D.}~\bibnamefont {Psaltis}},\ }\bibfield  {title} {\bibinfo {title} {Inference in artificial intelligence with deep optics and photonics},\ }\href {https://doi.org/10.1038/s41586-020-2973-6} {\bibfield  {journal} {\bibinfo  {journal} {Nature}\ }\textbf {\bibinfo {volume} {588}},\ \bibinfo {pages} {39} (\bibinfo {year} {2020})}\BibitemShut {NoStop}%
\bibitem [{\citenamefont {Romera}\ \emph {et~al.}(2018)\citenamefont {Romera}, \citenamefont {Talatchian}, \citenamefont {Tsunegi}, \citenamefont {Abreu~Araujo}, \citenamefont {Cros}, \citenamefont {Bortolotti}, \citenamefont {Trastoy}, \citenamefont {Yakushiji}, \citenamefont {Fukushima}, \citenamefont {Kubota}, \citenamefont {Yuasa}, \citenamefont {Ernoult}, \citenamefont {Vodenicarevic}, \citenamefont {Hirtzlin}, \citenamefont {Locatelli}, \citenamefont {Querlioz},\ and\ \citenamefont {Grollier}}]{romera_vowel_2018}%
  \BibitemOpen
  \bibfield  {author} {\bibinfo {author} {\bibfnamefont {M.}~\bibnamefont {Romera}}, \bibinfo {author} {\bibfnamefont {P.}~\bibnamefont {Talatchian}}, \bibinfo {author} {\bibfnamefont {S.}~\bibnamefont {Tsunegi}}, \bibinfo {author} {\bibfnamefont {F.}~\bibnamefont {Abreu~Araujo}}, \bibinfo {author} {\bibfnamefont {V.}~\bibnamefont {Cros}}, \bibinfo {author} {\bibfnamefont {P.}~\bibnamefont {Bortolotti}}, \bibinfo {author} {\bibfnamefont {J.}~\bibnamefont {Trastoy}}, \bibinfo {author} {\bibfnamefont {K.}~\bibnamefont {Yakushiji}}, \bibinfo {author} {\bibfnamefont {A.}~\bibnamefont {Fukushima}}, \bibinfo {author} {\bibfnamefont {H.}~\bibnamefont {Kubota}}, \bibinfo {author} {\bibfnamefont {S.}~\bibnamefont {Yuasa}}, \bibinfo {author} {\bibfnamefont {M.}~\bibnamefont {Ernoult}}, \bibinfo {author} {\bibfnamefont {D.}~\bibnamefont {Vodenicarevic}}, \bibinfo {author} {\bibfnamefont {T.}~\bibnamefont {Hirtzlin}}, \bibinfo {author} {\bibfnamefont {N.}~\bibnamefont {Locatelli}}, \bibinfo {author} {\bibfnamefont
  {D.}~\bibnamefont {Querlioz}},\ and\ \bibinfo {author} {\bibfnamefont {J.}~\bibnamefont {Grollier}},\ }\bibfield  {title} {\bibinfo {title} {Vowel recognition with four coupled spin-torque nano-oscillators},\ }\href {https://doi.org/10.1038/s41586-018-0632-y} {\bibfield  {journal} {\bibinfo  {journal} {Nature}\ }\textbf {\bibinfo {volume} {563}},\ \bibinfo {pages} {230} (\bibinfo {year} {2018})}\BibitemShut {NoStop}%
\bibitem [{\citenamefont {Chen}\ \emph {et~al.}(2023)\citenamefont {Chen}, \citenamefont {Nazhamaiti}, \citenamefont {Xu}, \citenamefont {Meng}, \citenamefont {Zhou}, \citenamefont {Li}, \citenamefont {Fan}, \citenamefont {Wei}, \citenamefont {Wu}, \citenamefont {Qiao}, \citenamefont {Fang},\ and\ \citenamefont {Dai}}]{chen_all-analog_2023}%
  \BibitemOpen
  \bibfield  {author} {\bibinfo {author} {\bibfnamefont {Y.}~\bibnamefont {Chen}}, \bibinfo {author} {\bibfnamefont {M.}~\bibnamefont {Nazhamaiti}}, \bibinfo {author} {\bibfnamefont {H.}~\bibnamefont {Xu}}, \bibinfo {author} {\bibfnamefont {Y.}~\bibnamefont {Meng}}, \bibinfo {author} {\bibfnamefont {T.}~\bibnamefont {Zhou}}, \bibinfo {author} {\bibfnamefont {G.}~\bibnamefont {Li}}, \bibinfo {author} {\bibfnamefont {J.}~\bibnamefont {Fan}}, \bibinfo {author} {\bibfnamefont {Q.}~\bibnamefont {Wei}}, \bibinfo {author} {\bibfnamefont {J.}~\bibnamefont {Wu}}, \bibinfo {author} {\bibfnamefont {F.}~\bibnamefont {Qiao}}, \bibinfo {author} {\bibfnamefont {L.}~\bibnamefont {Fang}},\ and\ \bibinfo {author} {\bibfnamefont {Q.}~\bibnamefont {Dai}},\ }\bibfield  {title} {\bibinfo {title} {All-analog photoelectronic chip for high-speed vision tasks},\ }\href {https://doi.org/10.1038/s41586-023-06558-8} {\bibfield  {journal} {\bibinfo  {journal} {Nature}\ }\textbf {\bibinfo {volume} {623}},\ \bibinfo {pages} {48} (\bibinfo
  {year} {2023})}\BibitemShut {NoStop}%
\bibitem [{\citenamefont {Lee}\ \emph {et~al.}(2022)\citenamefont {Lee}, \citenamefont {Mulder},\ and\ \citenamefont {Hopkins}}]{lee_mechanical_2022}%
  \BibitemOpen
  \bibfield  {author} {\bibinfo {author} {\bibfnamefont {R.~H.}\ \bibnamefont {Lee}}, \bibinfo {author} {\bibfnamefont {E.~A.~B.}\ \bibnamefont {Mulder}},\ and\ \bibinfo {author} {\bibfnamefont {J.~B.}\ \bibnamefont {Hopkins}},\ }\bibfield  {title} {\bibinfo {title} {Mechanical neural networks: Architected materials that learn behaviors},\ }\href {https://doi.org/10.1126/scirobotics.abq7278} {\bibfield  {journal} {\bibinfo  {journal} {Science Robotics}\ }\textbf {\bibinfo {volume} {7}},\ \bibinfo {pages} {eabq7278} (\bibinfo {year} {2022})}\BibitemShut {NoStop}%
\bibitem [{\citenamefont {Momeni}\ \emph {et~al.}(2023)\citenamefont {Momeni}, \citenamefont {Rahmani}, \citenamefont {Malléjac}, \citenamefont {del Hougne},\ and\ \citenamefont {Fleury}}]{science_backprop-free_prnn}%
  \BibitemOpen
  \bibfield  {author} {\bibinfo {author} {\bibfnamefont {A.}~\bibnamefont {Momeni}}, \bibinfo {author} {\bibfnamefont {B.}~\bibnamefont {Rahmani}}, \bibinfo {author} {\bibfnamefont {M.}~\bibnamefont {Malléjac}}, \bibinfo {author} {\bibfnamefont {P.}~\bibnamefont {del Hougne}},\ and\ \bibinfo {author} {\bibfnamefont {R.}~\bibnamefont {Fleury}},\ }\bibfield  {title} {\bibinfo {title} {Backpropagation-free training of deep physical neural networks},\ }\href {https://doi.org/10.1126/science.adi8474} {\bibfield  {journal} {\bibinfo  {journal} {Science}\ }\textbf {\bibinfo {volume} {382}},\ \bibinfo {pages} {1297} (\bibinfo {year} {2023})}\BibitemShut {NoStop}%
\bibitem [{\citenamefont {Stern}\ \emph {et~al.}(2021)\citenamefont {Stern}, \citenamefont {Hexner}, \citenamefont {Rocks},\ and\ \citenamefont {Liu}}]{Stern_2021}%
  \BibitemOpen
  \bibfield  {author} {\bibinfo {author} {\bibfnamefont {M.}~\bibnamefont {Stern}}, \bibinfo {author} {\bibfnamefont {D.}~\bibnamefont {Hexner}}, \bibinfo {author} {\bibfnamefont {J.~W.}\ \bibnamefont {Rocks}},\ and\ \bibinfo {author} {\bibfnamefont {A.~J.}\ \bibnamefont {Liu}},\ }\bibfield  {title} {\bibinfo {title} {Supervised learning in physical networks: From machine learning to learning machines},\ }\bibfield  {journal} {\bibinfo  {journal} {Physical Review X}\ }\textbf {\bibinfo {volume} {11}},\ \href {https://doi.org/10.1103/physrevx.11.021045} {10.1103/physrevx.11.021045} (\bibinfo {year} {2021})\BibitemShut {NoStop}%
\bibitem [{\citenamefont {Momeni}\ \emph {et~al.}(2024)\citenamefont {Momeni}, \citenamefont {Rahmani}, \citenamefont {Scellier}, \citenamefont {Wright}, \citenamefont {McMahon}, \citenamefont {Wanjura}, \citenamefont {Li}, \citenamefont {Skalli}, \citenamefont {Berloff}, \citenamefont {Onodera}, \citenamefont {Oguz}, \citenamefont {Morichetti}, \citenamefont {del Hougne}, \citenamefont {Gallo}, \citenamefont {Sebastian}, \citenamefont {Mirhoseini}, \citenamefont {Zhang}, \citenamefont {Marković}, \citenamefont {Brunner}, \citenamefont {Moser}, \citenamefont {Gigan}, \citenamefont {Marquardt}, \citenamefont {Ozcan}, \citenamefont {Grollier}, \citenamefont {Liu}, \citenamefont {Psaltis}, \citenamefont {Alù},\ and\ \citenamefont {Fleury}}]{momeni2024}%
  \BibitemOpen
  \bibfield  {author} {\bibinfo {author} {\bibfnamefont {A.}~\bibnamefont {Momeni}}, \bibinfo {author} {\bibfnamefont {B.}~\bibnamefont {Rahmani}}, \bibinfo {author} {\bibfnamefont {B.}~\bibnamefont {Scellier}}, \bibinfo {author} {\bibfnamefont {L.~G.}\ \bibnamefont {Wright}}, \bibinfo {author} {\bibfnamefont {P.~L.}\ \bibnamefont {McMahon}}, \bibinfo {author} {\bibfnamefont {C.~C.}\ \bibnamefont {Wanjura}}, \bibinfo {author} {\bibfnamefont {Y.}~\bibnamefont {Li}}, \bibinfo {author} {\bibfnamefont {A.}~\bibnamefont {Skalli}}, \bibinfo {author} {\bibfnamefont {N.~G.}\ \bibnamefont {Berloff}}, \bibinfo {author} {\bibfnamefont {T.}~\bibnamefont {Onodera}}, \bibinfo {author} {\bibfnamefont {I.}~\bibnamefont {Oguz}}, \bibinfo {author} {\bibfnamefont {F.}~\bibnamefont {Morichetti}}, \bibinfo {author} {\bibfnamefont {P.}~\bibnamefont {del Hougne}}, \bibinfo {author} {\bibfnamefont {M.~L.}\ \bibnamefont {Gallo}}, \bibinfo {author} {\bibfnamefont {A.}~\bibnamefont {Sebastian}}, \bibinfo {author} {\bibfnamefont
  {A.}~\bibnamefont {Mirhoseini}}, \bibinfo {author} {\bibfnamefont {C.}~\bibnamefont {Zhang}}, \bibinfo {author} {\bibfnamefont {D.}~\bibnamefont {Marković}}, \bibinfo {author} {\bibfnamefont {D.}~\bibnamefont {Brunner}}, \bibinfo {author} {\bibfnamefont {C.}~\bibnamefont {Moser}}, \bibinfo {author} {\bibfnamefont {S.}~\bibnamefont {Gigan}}, \bibinfo {author} {\bibfnamefont {F.}~\bibnamefont {Marquardt}}, \bibinfo {author} {\bibfnamefont {A.}~\bibnamefont {Ozcan}}, \bibinfo {author} {\bibfnamefont {J.}~\bibnamefont {Grollier}}, \bibinfo {author} {\bibfnamefont {A.~J.}\ \bibnamefont {Liu}}, \bibinfo {author} {\bibfnamefont {D.}~\bibnamefont {Psaltis}}, \bibinfo {author} {\bibfnamefont {A.}~\bibnamefont {Alù}},\ and\ \bibinfo {author} {\bibfnamefont {R.}~\bibnamefont {Fleury}},\ }\href {https://arxiv.org/abs/2406.03372} {\bibinfo {title} {Training of physical neural networks}} (\bibinfo {year} {2024}),\ \Eprint {https://arxiv.org/abs/2406.03372} {arXiv:2406.03372 [physics.app-ph]} \BibitemShut {NoStop}%
\bibitem [{\citenamefont {Rusch}\ and\ \citenamefont {Mishra}(2021)}]{rusch_coupled_2021}%
  \BibitemOpen
  \bibfield  {author} {\bibinfo {author} {\bibfnamefont {T.~K.}\ \bibnamefont {Rusch}}\ and\ \bibinfo {author} {\bibfnamefont {S.}~\bibnamefont {Mishra}},\ }\bibfield  {title} {\bibinfo {title} {Coupled {Oscillatory} {Recurrent} {Neural} {Network} ({coRNN}): {An} accurate and (gradient) stable architecture for learning long time dependencies},\ }\href {http://arxiv.org/abs/2010.00951} {\bibfield  {journal} {\bibinfo  {journal} {arXiv:2010.00951 [cs, stat]}\ } (\bibinfo {year} {2021})},\ \bibinfo {note} {arXiv: 2010.00951}\BibitemShut {NoStop}%
\bibitem [{\citenamefont {Effenberger}\ \emph {et~al.}(2025)\citenamefont {Effenberger}, \citenamefont {Carvalho}, \citenamefont {Dubinin},\ and\ \citenamefont {Singer}}]{Effenberger_doi:10.1073/pnas.2412830122}%
  \BibitemOpen
  \bibfield  {author} {\bibinfo {author} {\bibfnamefont {F.}~\bibnamefont {Effenberger}}, \bibinfo {author} {\bibfnamefont {P.}~\bibnamefont {Carvalho}}, \bibinfo {author} {\bibfnamefont {I.}~\bibnamefont {Dubinin}},\ and\ \bibinfo {author} {\bibfnamefont {W.}~\bibnamefont {Singer}},\ }\bibfield  {title} {\bibinfo {title} {The functional role of oscillatory dynamics in neocortical circuits: A computational perspective},\ }\href {https://doi.org/10.1073/pnas.2412830122} {\bibfield  {journal} {\bibinfo  {journal} {Proceedings of the National Academy of Sciences}\ }\textbf {\bibinfo {volume} {122}},\ \bibinfo {pages} {e2412830122} (\bibinfo {year} {2025})},\ \Eprint {https://arxiv.org/abs/https://www.pnas.org/doi/pdf/10.1073/pnas.2412830122} {https://www.pnas.org/doi/pdf/10.1073/pnas.2412830122} \BibitemShut {NoStop}%
\bibitem [{\citenamefont {Fries}(2015)}]{fries2015}%
  \BibitemOpen
  \bibfield  {author} {\bibinfo {author} {\bibfnamefont {P.}~\bibnamefont {Fries}},\ }\bibfield  {title} {\bibinfo {title} {Rhythms for cognition: Communication through coherence},\ }\href {https://doi.org/https://doi.org/10.1016/j.neuron.2015.09.034} {\bibfield  {journal} {\bibinfo  {journal} {Neuron}\ }\textbf {\bibinfo {volume} {88}},\ \bibinfo {pages} {220} (\bibinfo {year} {2015})}\BibitemShut {NoStop}%
\bibitem [{\citenamefont {Rosenblum}\ \emph {et~al.}(2001)\citenamefont {Rosenblum}, \citenamefont {Pikovsky}, \citenamefont {Kurths}, \citenamefont {Schäfer},\ and\ \citenamefont {Tass}}]{rosenblum_phase_2001}%
  \BibitemOpen
  \bibfield  {author} {\bibinfo {author} {\bibfnamefont {M.}~\bibnamefont {Rosenblum}}, \bibinfo {author} {\bibfnamefont {A.}~\bibnamefont {Pikovsky}}, \bibinfo {author} {\bibfnamefont {J.}~\bibnamefont {Kurths}}, \bibinfo {author} {\bibfnamefont {C.}~\bibnamefont {Schäfer}},\ and\ \bibinfo {author} {\bibfnamefont {P.}~\bibnamefont {Tass}},\ }\bibfield  {title} {\bibinfo {title} {Chapter 9 phase synchronization: From theory to data analysis},\ }in\ \href {https://doi.org/https://doi.org/10.1016/S1383-8121(01)80012-9} {\emph {\bibinfo {booktitle} {Handbook of Biological Physics}}},\ Vol.~\bibinfo {volume} {4},\ \bibinfo {editor} {edited by\ \bibinfo {editor} {\bibfnamefont {F.}~\bibnamefont {Moss}}\ and\ \bibinfo {editor} {\bibfnamefont {S.}~\bibnamefont {Gielen}}}\ (\bibinfo  {publisher} {North-Holland},\ \bibinfo {year} {2001})\ pp.\ \bibinfo {pages} {279--321}\BibitemShut {NoStop}%
\bibitem [{\citenamefont {Hughes}\ \emph {et~al.}(2019)\citenamefont {Hughes}, \citenamefont {Williamson}, \citenamefont {Minkov},\ and\ \citenamefont {Fan}}]{hughes2019}%
  \BibitemOpen
  \bibfield  {author} {\bibinfo {author} {\bibfnamefont {T.~W.}\ \bibnamefont {Hughes}}, \bibinfo {author} {\bibfnamefont {I.~A.~D.}\ \bibnamefont {Williamson}}, \bibinfo {author} {\bibfnamefont {M.}~\bibnamefont {Minkov}},\ and\ \bibinfo {author} {\bibfnamefont {S.}~\bibnamefont {Fan}},\ }\bibfield  {title} {\bibinfo {title} {Wave physics as an analog recurrent neural network},\ }\href {https://doi.org/10.1126/sciadv.aay6946} {\bibfield  {journal} {\bibinfo  {journal} {Science Advances}\ }\textbf {\bibinfo {volume} {5}},\ \bibinfo {pages} {eaay6946} (\bibinfo {year} {2019})}\BibitemShut {NoStop}%
\bibitem [{\citenamefont {Buzsáki}\ and\ \citenamefont {Wang}(2012)}]{buzsaki_mechanisms_2012}%
  \BibitemOpen
  \bibfield  {author} {\bibinfo {author} {\bibfnamefont {G.}~\bibnamefont {Buzsáki}}\ and\ \bibinfo {author} {\bibfnamefont {X.-J.}\ \bibnamefont {Wang}},\ }\bibfield  {title} {\bibinfo {title} {Mechanisms of {Gamma} {Oscillations}},\ }\href {https://doi.org/10.1146/annurev-neuro-062111-150444} {\bibfield  {journal} {\bibinfo  {journal} {Annual Review of Neuroscience}\ }\textbf {\bibinfo {volume} {35}},\ \bibinfo {pages} {203} (\bibinfo {year} {2012})}\BibitemShut {NoStop}%
\bibitem [{\citenamefont {Goltsev}\ \emph {et~al.}(2013)\citenamefont {Goltsev}, \citenamefont {Lopes}, \citenamefont {Lee},\ and\ \citenamefont {Mendes}}]{goltsev_critical_2013}%
  \BibitemOpen
  \bibfield  {author} {\bibinfo {author} {\bibfnamefont {A.~V.}\ \bibnamefont {Goltsev}}, \bibinfo {author} {\bibfnamefont {M.~A.}\ \bibnamefont {Lopes}}, \bibinfo {author} {\bibfnamefont {K.-E.}\ \bibnamefont {Lee}},\ and\ \bibinfo {author} {\bibfnamefont {J.~F.~F.}\ \bibnamefont {Mendes}},\ }\bibfield  {title} {\bibinfo {title} {Critical and resonance phenomena in neural networks},\ }\href {https://doi.org/10.1063/1.4776498} {\bibfield  {journal} {\bibinfo  {journal} {arXiv:1211.5686 [cond-mat, physics:physics, q-bio]}\ ,\ \bibinfo {pages} {28}} (\bibinfo {year} {2013})},\ \bibinfo {note} {arXiv: 1211.5686}\BibitemShut {NoStop}%
\bibitem [{\citenamefont {Dubinin}\ and\ \citenamefont {Effenberger}(2024)}]{DUBININ2024106179}%
  \BibitemOpen
  \bibfield  {author} {\bibinfo {author} {\bibfnamefont {I.}~\bibnamefont {Dubinin}}\ and\ \bibinfo {author} {\bibfnamefont {F.}~\bibnamefont {Effenberger}},\ }\bibfield  {title} {\bibinfo {title} {Fading memory as inductive bias in residual recurrent networks},\ }\href {https://doi.org/https://doi.org/10.1016/j.neunet.2024.106179} {\bibfield  {journal} {\bibinfo  {journal} {Neural Networks}\ }\textbf {\bibinfo {volume} {173}},\ \bibinfo {pages} {106179} (\bibinfo {year} {2024})}\BibitemShut {NoStop}%
\bibitem [{\citenamefont {Ashwin}\ and\ \citenamefont {Postlethwaite}(2021)}]{ashwin_excitable_2021}%
  \BibitemOpen
  \bibfield  {author} {\bibinfo {author} {\bibfnamefont {P.}~\bibnamefont {Ashwin}}\ and\ \bibinfo {author} {\bibfnamefont {C.}~\bibnamefont {Postlethwaite}},\ }\bibfield  {title} {\bibinfo {title} {Excitable networks for finite state computation with continuous time recurrent neural networks},\ }\href {https://doi.org/10.1007/s00422-021-00895-5} {\bibfield  {journal} {\bibinfo  {journal} {Biological Cybernetics}\ }\textbf {\bibinfo {volume} {115}},\ \bibinfo {pages} {519} (\bibinfo {year} {2021})}\BibitemShut {NoStop}%
\bibitem [{\citenamefont {Nikolić}\ \emph {et~al.}(2009)\citenamefont {Nikolić}, \citenamefont {Häusler}, \citenamefont {Singer},\ and\ \citenamefont {Maass}}]{Nikolic_distributed_2009}%
  \BibitemOpen
  \bibfield  {author} {\bibinfo {author} {\bibfnamefont {D.}~\bibnamefont {Nikolić}}, \bibinfo {author} {\bibfnamefont {S.}~\bibnamefont {Häusler}}, \bibinfo {author} {\bibfnamefont {W.}~\bibnamefont {Singer}},\ and\ \bibinfo {author} {\bibfnamefont {W.}~\bibnamefont {Maass}},\ }\bibfield  {title} {\bibinfo {title} {Distributed {Fading} {Memory} for {Stimulus} {Properties} in the {Primary} {Visual} {Cortex}},\ }\href {https://doi.org/10.1371/journal.pbio.1000260} {\bibfield  {journal} {\bibinfo  {journal} {PLoS Biology}\ }\textbf {\bibinfo {volume} {7}},\ \bibinfo {pages} {e1000260} (\bibinfo {year} {2009})}\BibitemShut {NoStop}%
\bibitem [{\citenamefont {Doelling}\ and\ \citenamefont {Poeppel}(2015)}]{Doelling_cortical_2015}%
  \BibitemOpen
  \bibfield  {author} {\bibinfo {author} {\bibfnamefont {K.~B.}\ \bibnamefont {Doelling}}\ and\ \bibinfo {author} {\bibfnamefont {D.}~\bibnamefont {Poeppel}},\ }\bibfield  {title} {\bibinfo {title} {Cortical entrainment to music and its modulation by expertise},\ }\bibfield  {journal} {\bibinfo  {journal} {Proceedings of the National Academy of Sciences}\ }\textbf {\bibinfo {volume} {112}},\ \href {https://doi.org/10.1073/pnas.1508431112} {10.1073/pnas.1508431112} (\bibinfo {year} {2015})\BibitemShut {NoStop}%
\bibitem [{\citenamefont {Schmidt}\ \emph {et~al.}(2023)\citenamefont {Schmidt}, \citenamefont {Rose},\ and\ \citenamefont {Muralidharan}}]{SCHMIDT2023102796}%
  \BibitemOpen
  \bibfield  {author} {\bibinfo {author} {\bibfnamefont {R.}~\bibnamefont {Schmidt}}, \bibinfo {author} {\bibfnamefont {J.}~\bibnamefont {Rose}},\ and\ \bibinfo {author} {\bibfnamefont {V.}~\bibnamefont {Muralidharan}},\ }\bibfield  {title} {\bibinfo {title} {Transient oscillations as computations for cognition: Analysis, modeling and function},\ }\href {https://doi.org/https://doi.org/10.1016/j.conb.2023.102796} {\bibfield  {journal} {\bibinfo  {journal} {Current Opinion in Neurobiology}\ }\textbf {\bibinfo {volume} {83}},\ \bibinfo {pages} {102796} (\bibinfo {year} {2023})}\BibitemShut {NoStop}%
\bibitem [{\citenamefont {Palmigiano}\ \emph {et~al.}(2017)\citenamefont {Palmigiano}, \citenamefont {Geisel}, \citenamefont {Wolf},\ and\ \citenamefont {Battaglia}}]{Palmigiano2017}%
  \BibitemOpen
  \bibfield  {author} {\bibinfo {author} {\bibfnamefont {A.}~\bibnamefont {Palmigiano}}, \bibinfo {author} {\bibfnamefont {T.}~\bibnamefont {Geisel}}, \bibinfo {author} {\bibfnamefont {F.}~\bibnamefont {Wolf}},\ and\ \bibinfo {author} {\bibfnamefont {D.}~\bibnamefont {Battaglia}},\ }\bibfield  {title} {\bibinfo {title} {Flexible information routing by transient synchrony},\ }\href {https://doi.org/10.1038/nn.4569} {\bibfield  {journal} {\bibinfo  {journal} {Nature Neuroscience}\ }\textbf {\bibinfo {volume} {20}},\ \bibinfo {pages} {1014} (\bibinfo {year} {2017})}\BibitemShut {NoStop}%
\bibitem [{\citenamefont {Helfrich}\ \emph {et~al.}(2018)\citenamefont {Helfrich}, \citenamefont {Fiebelkorn}, \citenamefont {Szczepanski}, \citenamefont {Lin}, \citenamefont {Parvizi}, \citenamefont {Knight},\ and\ \citenamefont {Kastner}}]{helfrich2018}%
  \BibitemOpen
  \bibfield  {author} {\bibinfo {author} {\bibfnamefont {R.~F.}\ \bibnamefont {Helfrich}}, \bibinfo {author} {\bibfnamefont {I.~C.}\ \bibnamefont {Fiebelkorn}}, \bibinfo {author} {\bibfnamefont {S.~M.}\ \bibnamefont {Szczepanski}}, \bibinfo {author} {\bibfnamefont {J.~J.}\ \bibnamefont {Lin}}, \bibinfo {author} {\bibfnamefont {J.}~\bibnamefont {Parvizi}}, \bibinfo {author} {\bibfnamefont {R.~T.}\ \bibnamefont {Knight}},\ and\ \bibinfo {author} {\bibfnamefont {S.}~\bibnamefont {Kastner}},\ }\bibfield  {title} {\bibinfo {title} {Neural mechanisms of sustained attention are rhythmic},\ }\href {https://doi.org/https://doi.org/10.1016/j.neuron.2018.07.032} {\bibfield  {journal} {\bibinfo  {journal} {Neuron}\ }\textbf {\bibinfo {volume} {99}},\ \bibinfo {pages} {854} (\bibinfo {year} {2018})}\BibitemShut {NoStop}%
\bibitem [{\citenamefont {Csaba}\ and\ \citenamefont {Porod}(2020)}]{Csaba2020}%
  \BibitemOpen
  \bibfield  {author} {\bibinfo {author} {\bibfnamefont {G.}~\bibnamefont {Csaba}}\ and\ \bibinfo {author} {\bibfnamefont {W.}~\bibnamefont {Porod}},\ }\bibfield  {title} {\bibinfo {title} {Coupled oscillators for computing: A review and perspective},\ }\href {https://doi.org/10.1063/1.5120412} {\bibfield  {journal} {\bibinfo  {journal} {Applied Physics Reviews}\ }\textbf {\bibinfo {volume} {7}},\ \bibinfo {pages} {011302} (\bibinfo {year} {2020})}\BibitemShut {NoStop}%
\bibitem [{\citenamefont {Ulmann}()}]{Model-1}%
  \BibitemOpen
  \bibfield  {author} {\bibinfo {author} {\bibfnamefont {B.}~\bibnamefont {Ulmann}},\ }\href {http://analogparadigm.com/downloads/handbook.pdf} {\emph {\bibinfo {title} {anabrid Model-1 Analog Computer User Manual}}},\ \bibinfo {note} {\texttt{http://analogparadigm.com/downloads/handbook.pdf}}\BibitemShut {NoStop}%
\bibitem [{\citenamefont {Chen}\ \emph {et~al.}(2025)\citenamefont {Chen}, \citenamefont {Yang}, \citenamefont {Chen}, \citenamefont {Wang}, \citenamefont {Wang}, \citenamefont {Wang}, \citenamefont {Tian}, \citenamefont {Yu}, \citenamefont {Chen}, \citenamefont {Lin}, \citenamefont {Zhu}, \citenamefont {He}, \citenamefont {Wu}, \citenamefont {Li}, \citenamefont {Zhang}, \citenamefont {Lin}, \citenamefont {Xu}, \citenamefont {Li}, \citenamefont {Zhang}, \citenamefont {Qi}, \citenamefont {Wang}, \citenamefont {Wang}, \citenamefont {Shang}, \citenamefont {Liu}, \citenamefont {Cheng},\ and\ \citenamefont {Liu}}]{Chen2025}%
  \BibitemOpen
  \bibfield  {author} {\bibinfo {author} {\bibfnamefont {H.}~\bibnamefont {Chen}}, \bibinfo {author} {\bibfnamefont {J.}~\bibnamefont {Yang}}, \bibinfo {author} {\bibfnamefont {J.}~\bibnamefont {Chen}}, \bibinfo {author} {\bibfnamefont {S.}~\bibnamefont {Wang}}, \bibinfo {author} {\bibfnamefont {S.}~\bibnamefont {Wang}}, \bibinfo {author} {\bibfnamefont {D.}~\bibnamefont {Wang}}, \bibinfo {author} {\bibfnamefont {X.}~\bibnamefont {Tian}}, \bibinfo {author} {\bibfnamefont {Y.}~\bibnamefont {Yu}}, \bibinfo {author} {\bibfnamefont {X.}~\bibnamefont {Chen}}, \bibinfo {author} {\bibfnamefont {Y.}~\bibnamefont {Lin}}, \bibinfo {author} {\bibfnamefont {Q.}~\bibnamefont {Zhu}}, \bibinfo {author} {\bibfnamefont {Y.}~\bibnamefont {He}}, \bibinfo {author} {\bibfnamefont {X.}~\bibnamefont {Wu}}, \bibinfo {author} {\bibfnamefont {Y.}~\bibnamefont {Li}}, \bibinfo {author} {\bibfnamefont {X.}~\bibnamefont {Zhang}}, \bibinfo {author} {\bibfnamefont {N.}~\bibnamefont {Lin}}, \bibinfo {author} {\bibfnamefont {M.}~\bibnamefont
  {Xu}}, \bibinfo {author} {\bibfnamefont {Y.}~\bibnamefont {Li}}, \bibinfo {author} {\bibfnamefont {X.}~\bibnamefont {Zhang}}, \bibinfo {author} {\bibfnamefont {X.}~\bibnamefont {Qi}}, \bibinfo {author} {\bibfnamefont {Z.}~\bibnamefont {Wang}}, \bibinfo {author} {\bibfnamefont {H.}~\bibnamefont {Wang}}, \bibinfo {author} {\bibfnamefont {D.}~\bibnamefont {Shang}}, \bibinfo {author} {\bibfnamefont {Q.}~\bibnamefont {Liu}}, \bibinfo {author} {\bibfnamefont {K.-T.}\ \bibnamefont {Cheng}},\ and\ \bibinfo {author} {\bibfnamefont {M.}~\bibnamefont {Liu}},\ }\bibfield  {title} {\bibinfo {title} {Continuous-time digital twin with analog memristive neural ordinary differential equation solver},\ }\href {https://doi.org/10.1126/sciadv.adr7571} {\bibfield  {journal} {\bibinfo  {journal} {Science Advances}\ }\textbf {\bibinfo {volume} {11}},\ \bibinfo {pages} {eadr7571} (\bibinfo {year} {2025})},\ \Eprint {https://arxiv.org/abs/https://www.science.org/doi/pdf/10.1126/sciadv.adr7571}
  {https://www.science.org/doi/pdf/10.1126/sciadv.adr7571} \BibitemShut {NoStop}%
\bibitem [{\citenamefont {LeCun}\ \emph {et~al.}(2010)\citenamefont {LeCun}, \citenamefont {Cortes},\ and\ \citenamefont {Burges}}]{lecun_mnist_2010}%
  \BibitemOpen
  \bibfield  {author} {\bibinfo {author} {\bibfnamefont {Y.}~\bibnamefont {LeCun}}, \bibinfo {author} {\bibfnamefont {C.}~\bibnamefont {Cortes}},\ and\ \bibinfo {author} {\bibfnamefont {C.}~\bibnamefont {Burges}},\ }\bibfield  {title} {\bibinfo {title} {{MNIST} handwritten digit database},\ }\href@noop {} {\bibfield  {journal} {\bibinfo  {journal} {ATT Labs [Online]. Available: http://yann.lecun.com/exdb/mnist}\ }\textbf {\bibinfo {volume} {2}} (\bibinfo {year} {2010})}\BibitemShut {NoStop}%
\bibitem [{\citenamefont {Tanaka}\ \emph {et~al.}(2019)\citenamefont {Tanaka}, \citenamefont {Yamane}, \citenamefont {Héroux}, \citenamefont {Nakane}, \citenamefont {Kanazawa}, \citenamefont {Takeda}, \citenamefont {Numata}, \citenamefont {Nakano},\ and\ \citenamefont {Hirose}}]{TANAKA2019100}%
  \BibitemOpen
  \bibfield  {author} {\bibinfo {author} {\bibfnamefont {G.}~\bibnamefont {Tanaka}}, \bibinfo {author} {\bibfnamefont {T.}~\bibnamefont {Yamane}}, \bibinfo {author} {\bibfnamefont {J.~B.}\ \bibnamefont {Héroux}}, \bibinfo {author} {\bibfnamefont {R.}~\bibnamefont {Nakane}}, \bibinfo {author} {\bibfnamefont {N.}~\bibnamefont {Kanazawa}}, \bibinfo {author} {\bibfnamefont {S.}~\bibnamefont {Takeda}}, \bibinfo {author} {\bibfnamefont {H.}~\bibnamefont {Numata}}, \bibinfo {author} {\bibfnamefont {D.}~\bibnamefont {Nakano}},\ and\ \bibinfo {author} {\bibfnamefont {A.}~\bibnamefont {Hirose}},\ }\bibfield  {title} {\bibinfo {title} {Recent advances in physical reservoir computing: A review},\ }\href {https://doi.org/https://doi.org/10.1016/j.neunet.2019.03.005} {\bibfield  {journal} {\bibinfo  {journal} {Neural Networks}\ }\textbf {\bibinfo {volume} {115}},\ \bibinfo {pages} {100} (\bibinfo {year} {2019})}\BibitemShut {NoStop}%
\bibitem [{\citenamefont {Du}\ \emph {et~al.}(2017)\citenamefont {Du}, \citenamefont {Cai}, \citenamefont {Zidan}, \citenamefont {Ma}, \citenamefont {Lee},\ and\ \citenamefont {Lu}}]{du2017}%
  \BibitemOpen
  \bibfield  {author} {\bibinfo {author} {\bibfnamefont {C.}~\bibnamefont {Du}}, \bibinfo {author} {\bibfnamefont {F.}~\bibnamefont {Cai}}, \bibinfo {author} {\bibfnamefont {M.~A.}\ \bibnamefont {Zidan}}, \bibinfo {author} {\bibfnamefont {W.}~\bibnamefont {Ma}}, \bibinfo {author} {\bibfnamefont {S.~H.}\ \bibnamefont {Lee}},\ and\ \bibinfo {author} {\bibfnamefont {W.~D.}\ \bibnamefont {Lu}},\ }\bibfield  {title} {\bibinfo {title} {Reservoir computing using dynamic memristors for temporal information processing},\ }\bibfield  {journal} {\bibinfo  {journal} {Nature Communications}\ }\textbf {\bibinfo {volume} {8}},\ \href {https://doi.org/10.1038/s41467-017-02337-y} {10.1038/s41467-017-02337-y} (\bibinfo {year} {2017})\BibitemShut {NoStop}%
\bibitem [{\citenamefont {Kneubühl}(1997)}]{kneubuhl_oscillations_1997}%
  \BibitemOpen
  \bibfield  {author} {\bibinfo {author} {\bibfnamefont {F.~K.}\ \bibnamefont {Kneubühl}},\ }\href {https://doi.org/10.1007/978-3-662-03468-2} {\emph {\bibinfo {title} {Oscillations and {Waves}}}}\ (\bibinfo  {publisher} {Springer Berlin Heidelberg},\ \bibinfo {address} {Berlin, Heidelberg},\ \bibinfo {year} {1997})\BibitemShut {NoStop}%
\bibitem [{\citenamefont {Hairer}\ \emph {et~al.}(2003)\citenamefont {Hairer}, \citenamefont {Lubich},\ and\ \citenamefont {Wanner}}]{hairer_geometric_2003}%
  \BibitemOpen
  \bibfield  {author} {\bibinfo {author} {\bibfnamefont {E.}~\bibnamefont {Hairer}}, \bibinfo {author} {\bibfnamefont {C.}~\bibnamefont {Lubich}},\ and\ \bibinfo {author} {\bibfnamefont {G.}~\bibnamefont {Wanner}},\ }\bibfield  {title} {\bibinfo {title} {Geometric numerical integration illustrated by the {Störmer}–{Verlet} method},\ }\href {https://doi.org/10.1017/S0962492902000144} {\bibfield  {journal} {\bibinfo  {journal} {Acta Numerica}\ }\textbf {\bibinfo {volume} {12}},\ \bibinfo {pages} {399} (\bibinfo {year} {2003})}\BibitemShut {NoStop}%
\bibitem [{\citenamefont {Baronig}\ \emph {et~al.}(2024{\natexlab{a}})\citenamefont {Baronig}, \citenamefont {Ferrand}, \citenamefont {Sabathiel},\ and\ \citenamefont {Legenstein}}]{legenstein_spatiotemporalprocessingspiking}%
  \BibitemOpen
  \bibfield  {author} {\bibinfo {author} {\bibfnamefont {M.}~\bibnamefont {Baronig}}, \bibinfo {author} {\bibfnamefont {R.}~\bibnamefont {Ferrand}}, \bibinfo {author} {\bibfnamefont {S.}~\bibnamefont {Sabathiel}},\ and\ \bibinfo {author} {\bibfnamefont {R.}~\bibnamefont {Legenstein}},\ }\href {https://arxiv.org/abs/2408.07517} {\bibinfo {title} {Advancing spatio-temporal processing in spiking neural networks through adaptation}} (\bibinfo {year} {2024}{\natexlab{a}}),\ \Eprint {https://arxiv.org/abs/2408.07517} {arXiv:2408.07517 [cs.NE]} \BibitemShut {NoStop}%
\bibitem [{\citenamefont {Paszke}\ \emph {et~al.}(2019)\citenamefont {Paszke}, \citenamefont {Gross}, \citenamefont {Massa}, \citenamefont {Lerer}, \citenamefont {Bradbury}, \citenamefont {Chanan}, \citenamefont {Killeen}, \citenamefont {Lin}, \citenamefont {Gimelshein}, \citenamefont {Antiga}, \citenamefont {Desmaison}, \citenamefont {Kopf}, \citenamefont {Yang}, \citenamefont {DeVito}, \citenamefont {Raison}, \citenamefont {Tejani}, \citenamefont {Chilamkurthy}, \citenamefont {Steiner}, \citenamefont {Fang}, \citenamefont {Bai},\ and\ \citenamefont {Chintala}}]{paszke_pytorch_2019}%
  \BibitemOpen
  \bibfield  {author} {\bibinfo {author} {\bibfnamefont {A.}~\bibnamefont {Paszke}}, \bibinfo {author} {\bibfnamefont {S.}~\bibnamefont {Gross}}, \bibinfo {author} {\bibfnamefont {F.}~\bibnamefont {Massa}}, \bibinfo {author} {\bibfnamefont {A.}~\bibnamefont {Lerer}}, \bibinfo {author} {\bibfnamefont {J.}~\bibnamefont {Bradbury}}, \bibinfo {author} {\bibfnamefont {G.}~\bibnamefont {Chanan}}, \bibinfo {author} {\bibfnamefont {T.}~\bibnamefont {Killeen}}, \bibinfo {author} {\bibfnamefont {Z.}~\bibnamefont {Lin}}, \bibinfo {author} {\bibfnamefont {N.}~\bibnamefont {Gimelshein}}, \bibinfo {author} {\bibfnamefont {L.}~\bibnamefont {Antiga}}, \bibinfo {author} {\bibfnamefont {A.}~\bibnamefont {Desmaison}}, \bibinfo {author} {\bibfnamefont {A.}~\bibnamefont {Kopf}}, \bibinfo {author} {\bibfnamefont {E.}~\bibnamefont {Yang}}, \bibinfo {author} {\bibfnamefont {Z.}~\bibnamefont {DeVito}}, \bibinfo {author} {\bibfnamefont {M.}~\bibnamefont {Raison}}, \bibinfo {author} {\bibfnamefont {A.}~\bibnamefont {Tejani}}, \bibinfo
  {author} {\bibfnamefont {S.}~\bibnamefont {Chilamkurthy}}, \bibinfo {author} {\bibfnamefont {B.}~\bibnamefont {Steiner}}, \bibinfo {author} {\bibfnamefont {L.}~\bibnamefont {Fang}}, \bibinfo {author} {\bibfnamefont {J.}~\bibnamefont {Bai}},\ and\ \bibinfo {author} {\bibfnamefont {S.}~\bibnamefont {Chintala}},\ }\bibfield  {title} {\bibinfo {title} {{PyTorch}: {An} {Imperative} {Style}, {High}-{Performance} {Deep} {Learning} {Library}},\ }in\ \href {http://papers.neurips.cc/paper/9015-pytorch-an-imperative-style-high-performance-deep-learning-library.pdf} {\emph {\bibinfo {booktitle} {Advances in {Neural} {Information} {Processing} {Systems} 32}}},\ \bibinfo {editor} {edited by\ \bibinfo {editor} {\bibfnamefont {H.}~\bibnamefont {Wallach}}, \bibinfo {editor} {\bibfnamefont {H.}~\bibnamefont {Larochelle}}, \bibinfo {editor} {\bibfnamefont {A.}~\bibnamefont {Beygelzimer}}, \bibinfo {editor} {\bibfnamefont {F.~d.}\ \bibnamefont {Alché-Buc}}, \bibinfo {editor} {\bibfnamefont {E.}~\bibnamefont {Fox}},\ and\
  \bibinfo {editor} {\bibfnamefont {R.}~\bibnamefont {Garnett}}}\ (\bibinfo  {publisher} {Curran Associates, Inc.},\ \bibinfo {year} {2019})\ pp.\ \bibinfo {pages} {8024--8035}\BibitemShut {NoStop}%
\bibitem [{ADA()}]{ADALM2000}%
  \BibitemOpen
  \href {https://www.analog.com/en/resources/evaluation-hardware-and-software/evaluation-boards-kits/adalm2000.html} {\bibinfo {title} {{ADALM2000} {W}ebsite}},\ \bibinfo {howpublished} {\url{https://www.analog.com/en/resources/evaluation-hardware-and-software/evaluation-boards-kits/adalm2000.html}}\BibitemShut {NoStop}%
\bibitem [{\citenamefont {Ulmann}\ and\ \citenamefont {Köppel}(2025)}]{anabrid2025teensylogger}%
  \BibitemOpen
  \bibfield  {author} {\bibinfo {author} {\bibfnamefont {B.}~\bibnamefont {Ulmann}}\ and\ \bibinfo {author} {\bibfnamefont {S.}~\bibnamefont {Köppel}},\ }\href@noop {} {\bibinfo {title} {{TeensyLogger: Open-source analog data logger based on Teensy}}},\ \bibinfo {howpublished} {\url{https://github.com/anabrid/TeensyLogger}} (\bibinfo {year} {2025})\BibitemShut {NoStop}%
\bibitem [{\citenamefont {Ulmann}(2022)}]{Ulmann_analog_computing_2022}%
  \BibitemOpen
  \bibfield  {author} {\bibinfo {author} {\bibfnamefont {B.}~\bibnamefont {Ulmann}},\ }\href {https://doi.org/doi:10.1515/9783110787740} {\emph {\bibinfo {title} {Analog Computing}}}\ (\bibinfo  {publisher} {De Gruyter Oldenbourg},\ \bibinfo {address} {Berlin, Boston},\ \bibinfo {year} {2022})\BibitemShut {NoStop}%
\bibitem [{\citenamefont {Ulmann}(2020)}]{Ulmann+2020}%
  \BibitemOpen
  \bibfield  {author} {\bibinfo {author} {\bibfnamefont {B.}~\bibnamefont {Ulmann}},\ }\href {https://doi.org/doi:10.1515/9783110662207} {\emph {\bibinfo {title} {Analog and Hybrid Computer Programming}}}\ (\bibinfo  {publisher} {De Gruyter Oldenbourg},\ \bibinfo {address} {Berlin, Boston},\ \bibinfo {year} {2020})\BibitemShut {NoStop}%
\bibitem [{\citenamefont {Ohkubo}\ and\ \citenamefont {Inubushi}(2024)}]{Ohkubo2024ReservoirComputing}%
  \BibitemOpen
  \bibfield  {author} {\bibinfo {author} {\bibfnamefont {A.}~\bibnamefont {Ohkubo}}\ and\ \bibinfo {author} {\bibfnamefont {M.}~\bibnamefont {Inubushi}},\ }\bibfield  {title} {\bibinfo {title} {Reservoir computing with generalized readout based on generalized synchronization},\ }\href {https://doi.org/10.1038/s41598-024-81880-3} {\bibfield  {journal} {\bibinfo  {journal} {Scientific Reports}\ }\textbf {\bibinfo {volume} {14}},\ \bibinfo {pages} {30918} (\bibinfo {year} {2024})}\BibitemShut {NoStop}%
\bibitem [{\citenamefont {Satyanarayanan}(2017)}]{Satyanarayanan2017}%
  \BibitemOpen
  \bibfield  {author} {\bibinfo {author} {\bibfnamefont {M.}~\bibnamefont {Satyanarayanan}},\ }\bibfield  {title} {\bibinfo {title} {The emergence of edge computing},\ }\href {https://doi.org/10.1109/MC.2017.9} {\bibfield  {journal} {\bibinfo  {journal} {Computer}\ }\textbf {\bibinfo {volume} {50}},\ \bibinfo {pages} {30} (\bibinfo {year} {2017})}\BibitemShut {NoStop}%
\bibitem [{\citenamefont {Amit}(1989)}]{Amit_1989}%
  \BibitemOpen
  \bibfield  {author} {\bibinfo {author} {\bibfnamefont {D.~J.}\ \bibnamefont {Amit}},\ }\href@noop {} {\emph {\bibinfo {title} {Modeling Brain Function: The World of Attractor Neural Networks}}}\ (\bibinfo  {publisher} {Cambridge University Press},\ \bibinfo {year} {1989})\BibitemShut {NoStop}%
\bibitem [{\citenamefont {Indiveri}\ and\ \citenamefont {Liu}(2015)}]{indiveri2015}%
  \BibitemOpen
  \bibfield  {author} {\bibinfo {author} {\bibfnamefont {G.}~\bibnamefont {Indiveri}}\ and\ \bibinfo {author} {\bibfnamefont {S.-C.}\ \bibnamefont {Liu}},\ }\bibfield  {title} {\bibinfo {title} {Memory and information processing in neuromorphic systems},\ }\href {https://doi.org/10.1109/JPROC.2015.2444094} {\bibfield  {journal} {\bibinfo  {journal} {Proceedings of the IEEE}\ }\textbf {\bibinfo {volume} {103}},\ \bibinfo {pages} {1379} (\bibinfo {year} {2015})}\BibitemShut {NoStop}%
\bibitem [{\citenamefont {Kendall}\ \emph {et~al.}(2020)\citenamefont {Kendall}, \citenamefont {Pantone}, \citenamefont {Manickavasagam}, \citenamefont {Bengio},\ and\ \citenamefont {Scellier}}]{Kendall2020}%
  \BibitemOpen
  \bibfield  {author} {\bibinfo {author} {\bibfnamefont {J.~D.}\ \bibnamefont {Kendall}}, \bibinfo {author} {\bibfnamefont {R.~D.}\ \bibnamefont {Pantone}}, \bibinfo {author} {\bibfnamefont {K.}~\bibnamefont {Manickavasagam}}, \bibinfo {author} {\bibfnamefont {Y.}~\bibnamefont {Bengio}},\ and\ \bibinfo {author} {\bibfnamefont {B.}~\bibnamefont {Scellier}},\ }\bibfield  {title} {\bibinfo {title} {Training end-to-end analog neural networks with equilibrium propagation},\ }\href {https://arxiv.org/abs/2006.01981} {\bibfield  {journal} {\bibinfo  {journal} {CoRR}\ }\textbf {\bibinfo {volume} {abs/2006.01981}} (\bibinfo {year} {2020})},\ \Eprint {https://arxiv.org/abs/2006.01981} {2006.01981} \BibitemShut {NoStop}%
\bibitem [{\citenamefont {Kudithipudi}\ \emph {et~al.}(2025)\citenamefont {Kudithipudi}, \citenamefont {Schuman}, \citenamefont {Vineyard}, \citenamefont {Pandit}, \citenamefont {Merkel}, \citenamefont {Kubendran}, \citenamefont {Aimone}, \citenamefont {Orchard}, \citenamefont {Mayr}, \citenamefont {Benosman}, \citenamefont {Hays}, \citenamefont {Young}, \citenamefont {Bartolozzi}, \citenamefont {Majumdar}, \citenamefont {Cardwell}, \citenamefont {Payvand}, \citenamefont {Buckley}, \citenamefont {Kulkarni}, \citenamefont {Gonzalez}, \citenamefont {Cauwenberghs}, \citenamefont {Thakur}, \citenamefont {Subramoney},\ and\ \citenamefont {Furber}}]{Kudithipudi2025}%
  \BibitemOpen
  \bibfield  {author} {\bibinfo {author} {\bibfnamefont {D.}~\bibnamefont {Kudithipudi}}, \bibinfo {author} {\bibfnamefont {C.}~\bibnamefont {Schuman}}, \bibinfo {author} {\bibfnamefont {C.~M.}\ \bibnamefont {Vineyard}}, \bibinfo {author} {\bibfnamefont {T.}~\bibnamefont {Pandit}}, \bibinfo {author} {\bibfnamefont {C.}~\bibnamefont {Merkel}}, \bibinfo {author} {\bibfnamefont {R.}~\bibnamefont {Kubendran}}, \bibinfo {author} {\bibfnamefont {J.~B.}\ \bibnamefont {Aimone}}, \bibinfo {author} {\bibfnamefont {G.}~\bibnamefont {Orchard}}, \bibinfo {author} {\bibfnamefont {C.}~\bibnamefont {Mayr}}, \bibinfo {author} {\bibfnamefont {R.}~\bibnamefont {Benosman}}, \bibinfo {author} {\bibfnamefont {J.}~\bibnamefont {Hays}}, \bibinfo {author} {\bibfnamefont {C.}~\bibnamefont {Young}}, \bibinfo {author} {\bibfnamefont {C.}~\bibnamefont {Bartolozzi}}, \bibinfo {author} {\bibfnamefont {A.}~\bibnamefont {Majumdar}}, \bibinfo {author} {\bibfnamefont {S.~G.}\ \bibnamefont {Cardwell}}, \bibinfo {author} {\bibfnamefont
  {M.}~\bibnamefont {Payvand}}, \bibinfo {author} {\bibfnamefont {S.}~\bibnamefont {Buckley}}, \bibinfo {author} {\bibfnamefont {S.}~\bibnamefont {Kulkarni}}, \bibinfo {author} {\bibfnamefont {H.~A.}\ \bibnamefont {Gonzalez}}, \bibinfo {author} {\bibfnamefont {G.}~\bibnamefont {Cauwenberghs}}, \bibinfo {author} {\bibfnamefont {C.~S.}\ \bibnamefont {Thakur}}, \bibinfo {author} {\bibfnamefont {A.}~\bibnamefont {Subramoney}},\ and\ \bibinfo {author} {\bibfnamefont {S.}~\bibnamefont {Furber}},\ }\bibfield  {title} {\bibinfo {title} {Neuromorphic computing at scale},\ }\href {https://doi.org/10.1038/s41586-024-08253-8} {\bibfield  {journal} {\bibinfo  {journal} {Nature}\ }\textbf {\bibinfo {volume} {637}},\ \bibinfo {pages} {801} (\bibinfo {year} {2025})}\BibitemShut {NoStop}%
\bibitem [{\citenamefont {Baronig}\ \emph {et~al.}(2024{\natexlab{b}})\citenamefont {Baronig}, \citenamefont {Ferrand}, \citenamefont {Sabathiel},\ and\ \citenamefont {Legenstein}}]{Baronig2024AdvancingSpatioTemporal}%
  \BibitemOpen
  \bibfield  {author} {\bibinfo {author} {\bibfnamefont {M.}~\bibnamefont {Baronig}}, \bibinfo {author} {\bibfnamefont {R.}~\bibnamefont {Ferrand}}, \bibinfo {author} {\bibfnamefont {S.}~\bibnamefont {Sabathiel}},\ and\ \bibinfo {author} {\bibfnamefont {R.}~\bibnamefont {Legenstein}},\ }\href {https://doi.org/10.48550/arXiv.2408.07517} {\bibinfo {title} {Advancing {{Spatio-Temporal Processing}} in {{Spiking Neural Networks}} through {{Adaptation}}}} (\bibinfo {year} {2024}{\natexlab{b}})\BibitemShut {NoStop}%
\bibitem [{\citenamefont {Higuchi}\ \emph {et~al.}(2024)\citenamefont {Higuchi}, \citenamefont {Kairat}, \citenamefont {Boht{\'{e}}},\ and\ \citenamefont {Otte}}]{higuchi2024balanced}%
  \BibitemOpen
  \bibfield  {author} {\bibinfo {author} {\bibfnamefont {S.}~\bibnamefont {Higuchi}}, \bibinfo {author} {\bibfnamefont {S.}~\bibnamefont {Kairat}}, \bibinfo {author} {\bibfnamefont {S.~M.}\ \bibnamefont {Boht{\'{e}}}},\ and\ \bibinfo {author} {\bibfnamefont {S.}~\bibnamefont {Otte}},\ }\bibfield  {title} {\bibinfo {title} {{B}alanced {R}esonate-and-{F}ire {N}eurons},\ }in\ \href {https://openreview.net/forum?id=dkdilv4XD4} {\emph {\bibinfo {booktitle} {Proc. {ICML} 2024}}}\ (\bibinfo {year} {2024})\BibitemShut {NoStop}%
\bibitem [{\citenamefont {Pedregosa}\ \emph {et~al.}(2011)\citenamefont {Pedregosa}, \citenamefont {Varoquaux}, \citenamefont {Gramfort}, \citenamefont {Michel}, \citenamefont {Thirion}, \citenamefont {Grisel}, \citenamefont {Blondel}, \citenamefont {Prettenhofer}, \citenamefont {Weiss}, \citenamefont {Dubourg}, \citenamefont {Vanderplas}, \citenamefont {Passos}, \citenamefont {Cournapeau}, \citenamefont {Brucher}, \citenamefont {Perrot},\ and\ \citenamefont {Duchesnay}}]{scikit-learn}%
  \BibitemOpen
  \bibfield  {author} {\bibinfo {author} {\bibfnamefont {F.}~\bibnamefont {Pedregosa}}, \bibinfo {author} {\bibfnamefont {G.}~\bibnamefont {Varoquaux}}, \bibinfo {author} {\bibfnamefont {A.}~\bibnamefont {Gramfort}}, \bibinfo {author} {\bibfnamefont {V.}~\bibnamefont {Michel}}, \bibinfo {author} {\bibfnamefont {B.}~\bibnamefont {Thirion}}, \bibinfo {author} {\bibfnamefont {O.}~\bibnamefont {Grisel}}, \bibinfo {author} {\bibfnamefont {M.}~\bibnamefont {Blondel}}, \bibinfo {author} {\bibfnamefont {P.}~\bibnamefont {Prettenhofer}}, \bibinfo {author} {\bibfnamefont {R.}~\bibnamefont {Weiss}}, \bibinfo {author} {\bibfnamefont {V.}~\bibnamefont {Dubourg}}, \bibinfo {author} {\bibfnamefont {J.}~\bibnamefont {Vanderplas}}, \bibinfo {author} {\bibfnamefont {A.}~\bibnamefont {Passos}}, \bibinfo {author} {\bibfnamefont {D.}~\bibnamefont {Cournapeau}}, \bibinfo {author} {\bibfnamefont {M.}~\bibnamefont {Brucher}}, \bibinfo {author} {\bibfnamefont {M.}~\bibnamefont {Perrot}},\ and\ \bibinfo {author} {\bibfnamefont
  {E.}~\bibnamefont {Duchesnay}},\ }\bibfield  {title} {\bibinfo {title} {Scikit-learn: Machine learning in {P}ython},\ }\href@noop {} {\bibfield  {journal} {\bibinfo  {journal} {Journal of Machine Learning Research}\ }\textbf {\bibinfo {volume} {12}},\ \bibinfo {pages} {2825} (\bibinfo {year} {2011})}\BibitemShut {NoStop}%
\end{thebibliography}%

\appendix

\section{Error Metrics}\label{apx:error}

\begin{figure*}
    \includegraphics[width=\linewidth]{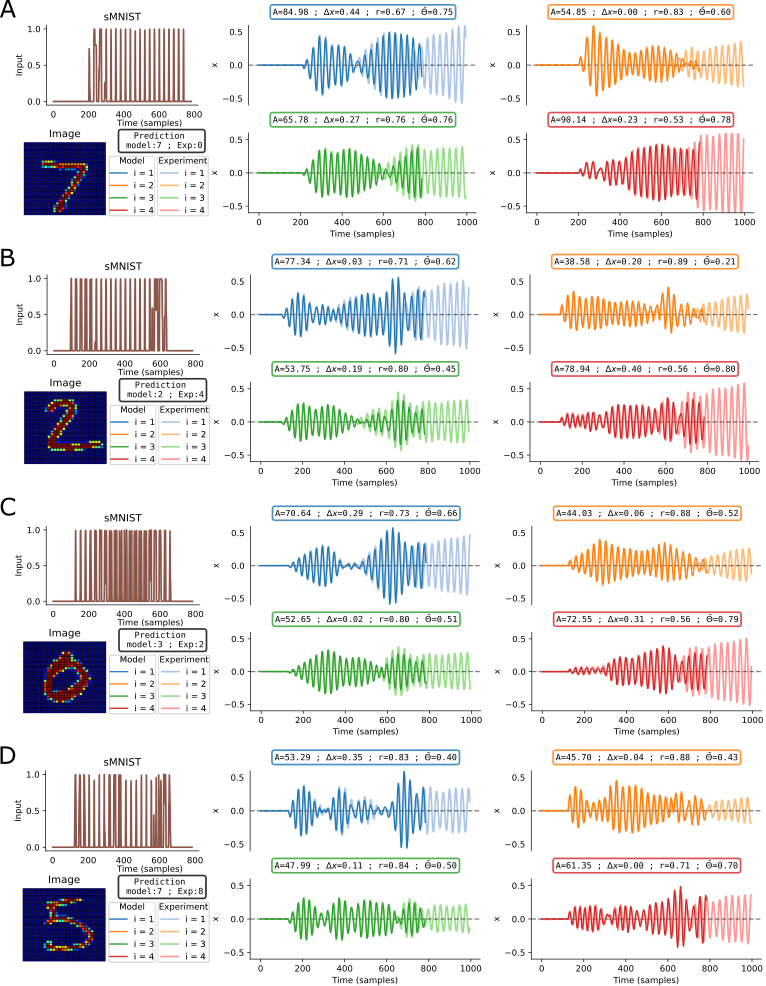}
    \caption{Additional examples of the analog and digital dynamics comparison across
    different input samples.
    Left column displays input in both sMNIST (serialized) and MNIST (image)
    formats.
    Center and right columns show the corresponding node dynamics, with light
    traces representing analog implementation and dark traces representing
    digital twin dynamics.}
    \label{fig:Network_samples}
\end{figure*}

To assess the difference between the dynamics of the analog implementation and
its digital counterpart, we defined the following error metrics, calculated
node-wise between the analog implementation and the digital twin:
\begin{itemize}
\item \textbf{Mismatch}: The difference in amplitudes at the last time step.
\item \textbf{Area}: The total area between the time series.
\item \textbf{Phase}: The average instantaneous phase difference.
\item \textbf{Correlation}: The correlation between the two time series.
\end{itemize}
These metrics can be visualized in Fig.~\ref{fig:4}F, and network samples
(Fig.~\ref{fig:Network_samples}) provide an intuitive qualitative understanding
of how these values relate to dynamics reproduction.

\begin{figure*}
    \includegraphics[width=\linewidth]{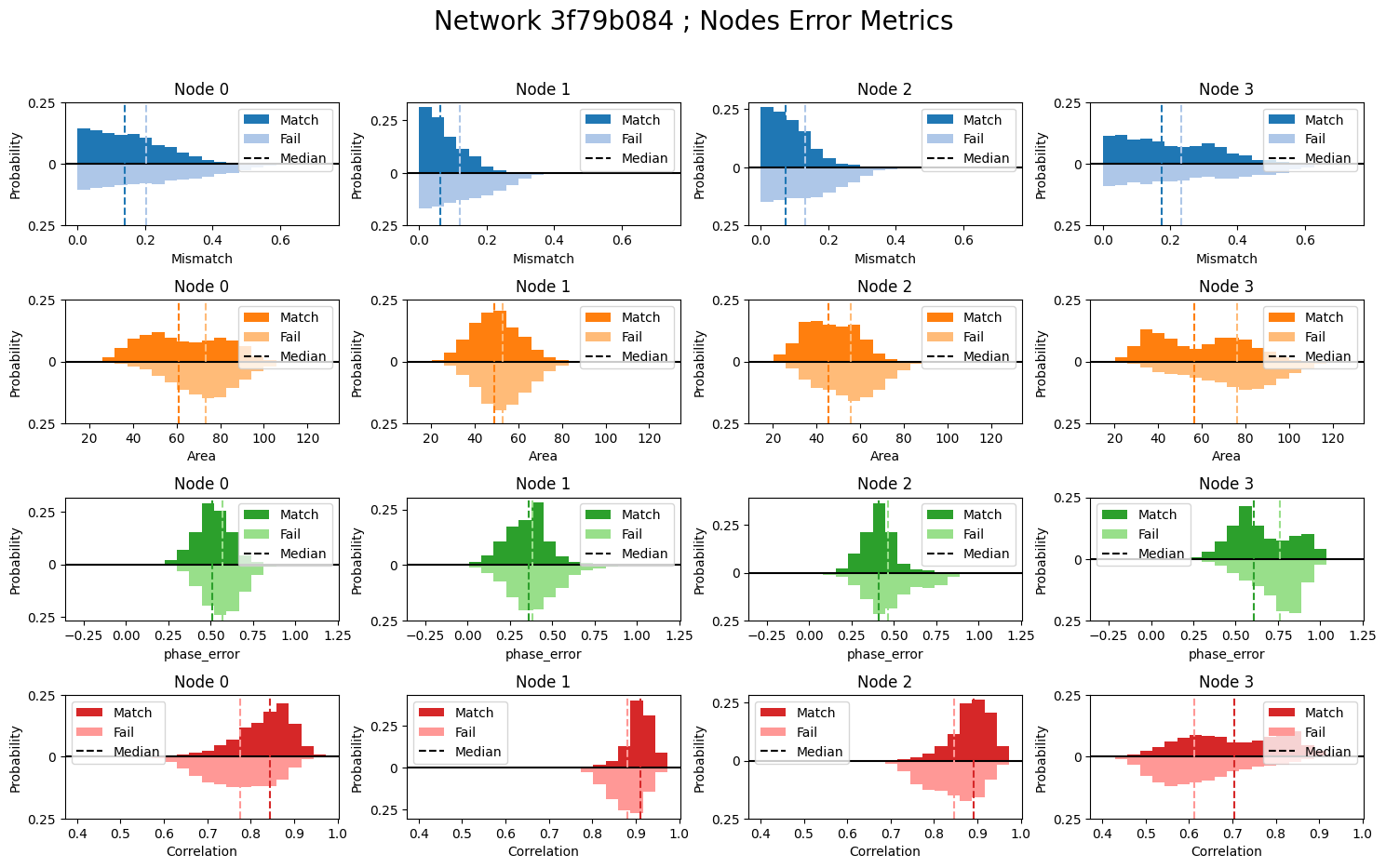}
    \caption{Node-wise error metric distributions for prediction agreement
    analysis.
    Probability densities of error metrics (mismatch, area, phase, correlation)
    for each of the four nodes, separated by cases where analog and digital
    predictions match versus fail to match.}
    \label{fig:Error_samples}
\end{figure*}

Comparative histograms of these error metrics for cases where labels match
versus fail to match between analog and digital implementations are shown for
each node individually in Fig.~\ref{fig:Error_samples}.
The concatenated data are presented in a condensed format in Fig.~\ref{fig:4}~D.

\section{Scaling Algorithm}\label{apx:Scaling_algorithm}

\begin{figure*}
    \centering
    \includegraphics[width=\linewidth]{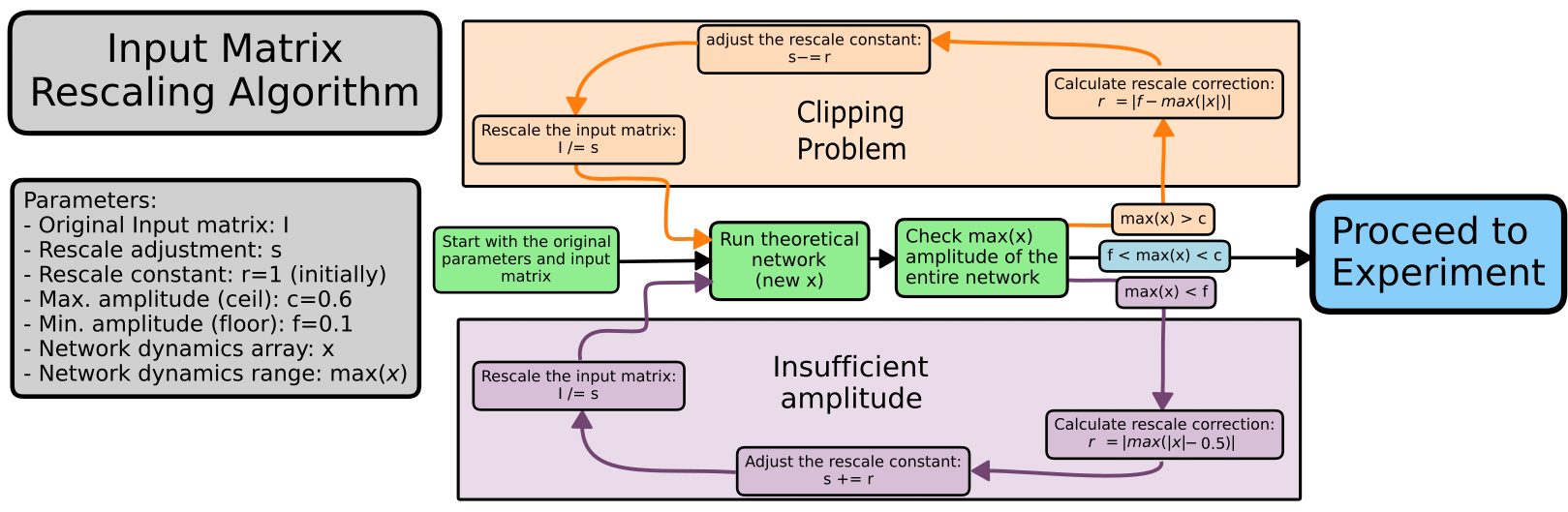}
    \caption{Flowchart of the input matrix rescaling algorithm.
    This iterative procedure optimizes the dynamic range usage of the analog implementation, avoiding both clipping and machine precision artifacts.}
    \label{fig:rescale-scheme}
\end{figure*}

Prior to any run of the analog implementation, the digital computer
(Fig.~\ref{fig:1}~A, item 1) loads the network parameters and the input sample.
With the network parameters, the digital computer undergoes a simulation loop
(Fig.~\ref{fig:rescale-scheme}) to obtain a rescaling factor $s$.
This rescaling factor is applied to the input matrix $I$ (Eq.~\ref{eq:forcing})
to ensure that the network dynamic range can fit into the analog implementation
non-clipping amplitude interval.
To illustrate the rescaling algorithm operation (Fig.~\ref{fig:rescale-scheme}),
consider the case where the digital computer loads the network data and
simulates the model for the first time.

After simulating the network for the first time, the digital computer obtains
the network dynamic range ($\max(|x|)$).
This value should lie within a pre-defined range $[floor,ceil]$.
If $\max(|x|)<floor$, the amplitude is insufficient, meaning that the network
dynamic range is comparable to the machine precision limit.
A rescaling correction, $r$, is then calculated to alter the scaling factor $s$.
In the case of insufficient amplitude, $s$ is increased and a new simulation is
run.
If $\max(|x|)>ceil$ the experiment might face clipping problems; in this case
$s$ is reduced and a new simulation is run.
This process is repeated until $\max(|x|)$ lies inside the desired experiment
range (Fig.~\ref{fig:rescale-scheme}).

% \paragraph{Pseudocode.}
% Let $s$ be the input scaling factor, with bounds $s_{\min}$ and $s_{\max}$, and
% target amplitude window $[\texttt{floor}, \texttt{ceil}]$.
% At each iteration, simulate the digital model with scaled inputs $I' = s I$ and
% compute $a = \max_t \max_i |x_i(t)|$.
% If $a < \texttt{floor}$, set $s \leftarrow \min(\beta_\uparrow s, s_{\max})$;
% if $a > \texttt{ceil}$, set $s \leftarrow \max(s/\beta_\downarrow, s_{\min})$.
% Terminate when $|a - a^*| < \varepsilon$ with $a^* \in [\texttt{floor}, \texttt{ceil}]$ or
% when reaching a maximum number of iterations.
% In our experiments we used $\texttt{floor}=0.1$, $\texttt{ceil}=0.6$, $\beta_\uparrow=1.1$,
% $\beta_\downarrow=1.1$, $\varepsilon=10^{-3}$, and a cap of 50 iterations.

Since resonance effects are expected, we opted to set $\texttt{ceil} = 0.6$,
leaving a margin for the system before reaching its limits, and set $floor =
0.1$.

\section{Original decoder}\label{apx:orig_decoder}

The straightforward approach for decoding the analog circuit output uses the
same decoder as the digital twin: an affine readout layer with readout weights
determined by the digital BPTT training procedure.

\begin{figure*}
    \centering
    \includegraphics[width=0.5\linewidth]{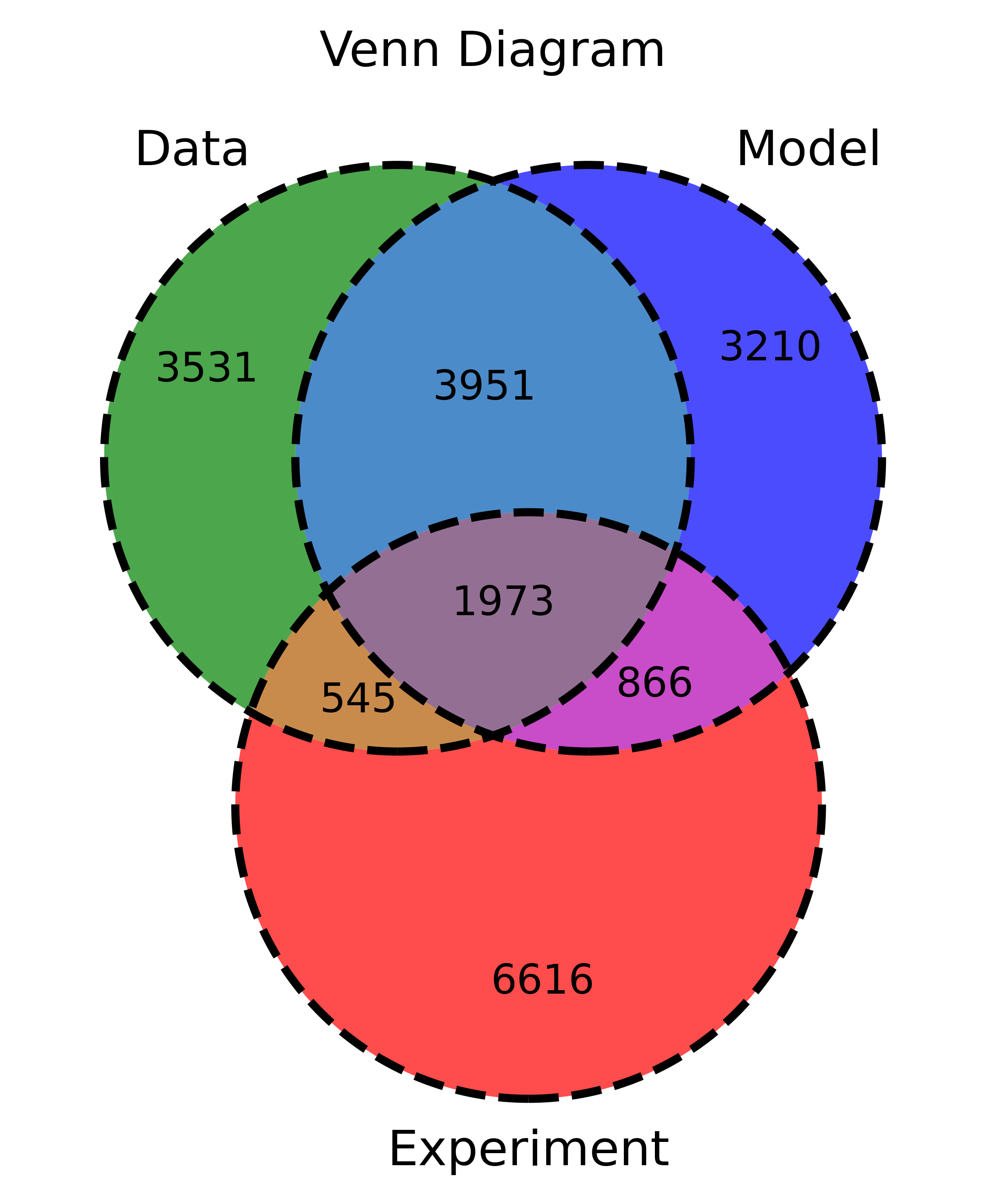}
    \caption{Prediction agreement analysis using Venn diagram representation.
    Overlap regions show the intersection between predictions from the analog
    implementation, digital twin, and true MNIST labels across the test
    dataset.}
    \label{fig:apx_venn}
\end{figure*}

The decoding performance can be visualized using a Venn diagram showing the
coincidence of the ground-truth labels, the label predictions by the digital
twin, and the label predictions by the analog twin on a per-class basis
(Fig.~\ref{fig:apx_venn}).

\begin{figure*}
    \includegraphics[width=\linewidth]{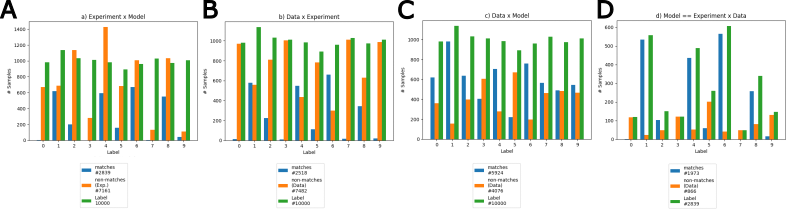}
    \caption{Digit-wise classification performance comparison across different
    evaluation scenarios.
    \textbf{A.} Analog implementation performance versus true dataset labels.
    \textbf{B.} Digital twin performance versus true dataset labels.
    \textbf{C.} Agreement between analog implementation and digital twin
    predictions.
    \textbf{D.} Accuracy of the analog-digital agreement subset when compared to
    true labels, showing how well cases where both systems agree actually
    correspond to correct classifications.}
    \label{fig:original_decoder_3f79b084}
\end{figure*}

We next examine whether decoding performance is consistent across all digit classes or if certain digits are decoded with superior or inferior accuracy (Fig.~\ref{fig:original_decoder_3f79b084}). 
The analog twin performs best for the digit classes 1, 4, 6, and 8.


\section{SVM decoding}\label{apx:svm_decoder}

An alternative method for assessing the networks' performance involves utilizing
the analog HORN as a reservoir.
In this approach, an independent readout (linear SVM) is trained on the output
of a pre-trained network while keeping the network parameters fixed.
The linear SVM was implemented using the scikit learn python package~\cite{scikit-learn}.

\section{Velocity-Coupled Network}\label{apx:velocity-network}
Here, we present the obtained results for the analysis of the physical harmonic
oscillator recurrent network with its nodes velocity-coupled.

In this network, the nodes are coupled through their velocity terms. This coupling leads to a modified form of the update equations (Eq.~\ref{eq:update_equations}), expressed as follows:

\begin{equation}\label{eq:y-coupled_update_equations}
\begin{split}
    x_{i, t+1} &= x_{i, t} + h y_{i, t+1},\\
    y_{i, t+1} &= y_{i, t} + h\left[  \sum^{n}_{j \neq i}{\left(W_{ji}y_{i, t}\right)} + I_{i}s(t) - 2\gamma y_{i, t}  - \omega_{i}^{2} x_{i, t} \right].
\end{split}
\end{equation}

For the readout, we used a "Hilbert decoder".
The Hilbert decoder computes the analytic signal by applying the discrete Hilbert
transform to each node's amplitude time series $x_i(t)$ to obtain the complex-valued analytic signal $\phi_i(t)$ on a time window $\delta T$ (default window length 100 time steps) around the readout time $t^*$.
We then construct an 8-dimensional feature vector by splitting $\phi_i(t^*)$ into its real and complex parts.
This feature vector is read out by an affine readout layer as previously to perform a digit class prediction.
This readout exploits phase information that is directly available in the
velocity-coupled network.

\begin{figure}
    \centering
    \includegraphics[width=\linewidth]{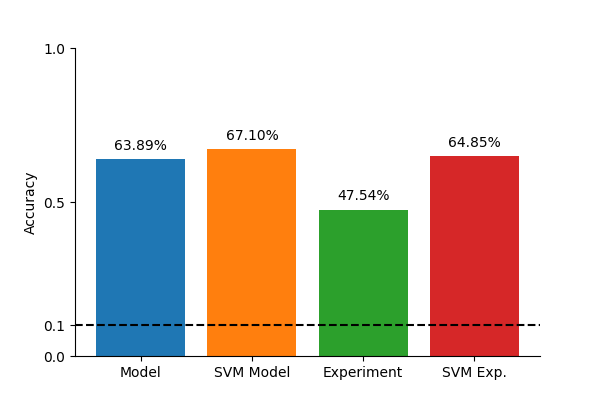}
    \caption{Classification performance comparison for velocity-coupled HORN
    implementation. Shows accuracy metrics for different readout strategies
    applied to the velocity-coupled network variant.}
    \label{fig:Y-Net_performance}
\end{figure}

\begin{figure*}
    \centering
    \includegraphics[width=\linewidth]{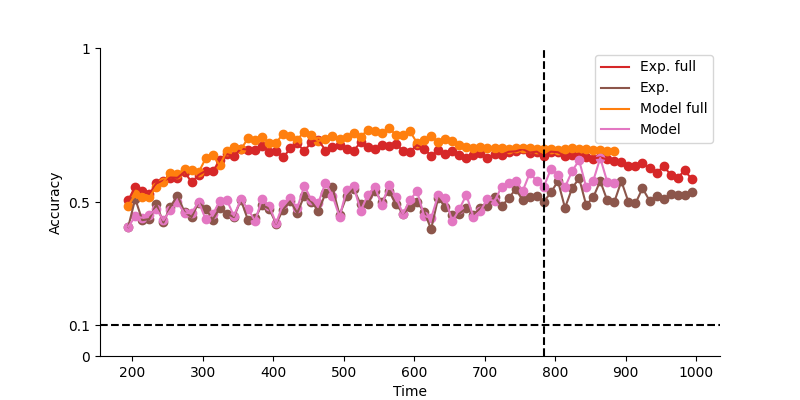}
    \caption{SVM readout performance analysis for velocity-coupled network.
    Temporal evolution of classification accuracy using both amplitude and
    velocity variables for decoding.}
    \label{fig:Y-Net_SVM_timeseries}
\end{figure*}

\begin{figure}
    \centering
    \includegraphics[width=\linewidth]{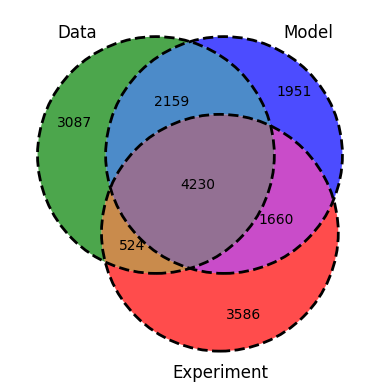}
    \caption{Prediction agreement analysis for velocity-coupled network
    implementation.
    Venn diagram showing overlaps between analog implementation, digital twin,
    and true label predictions.}
    \label{fig:Y-Net_Venn}
\end{figure}

\begin{figure*}
    \centering
    \includegraphics[width=\textwidth]{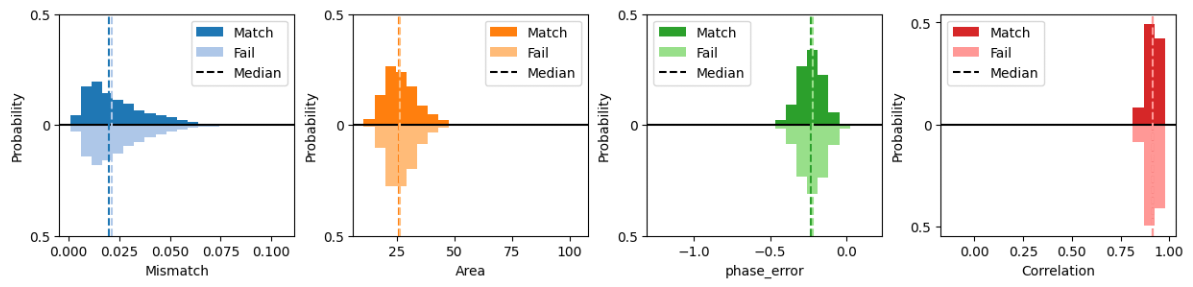}
    \caption{Error metric distributions for velocity-coupled network.
    Comparison of error metrics between analog implementation and digital twin
    for the velocity-coupled variant.}
    \label{fig:Y-Net_errors}
\end{figure*}

\begin{figure*}
    \centering
    \includegraphics[width=\textwidth]{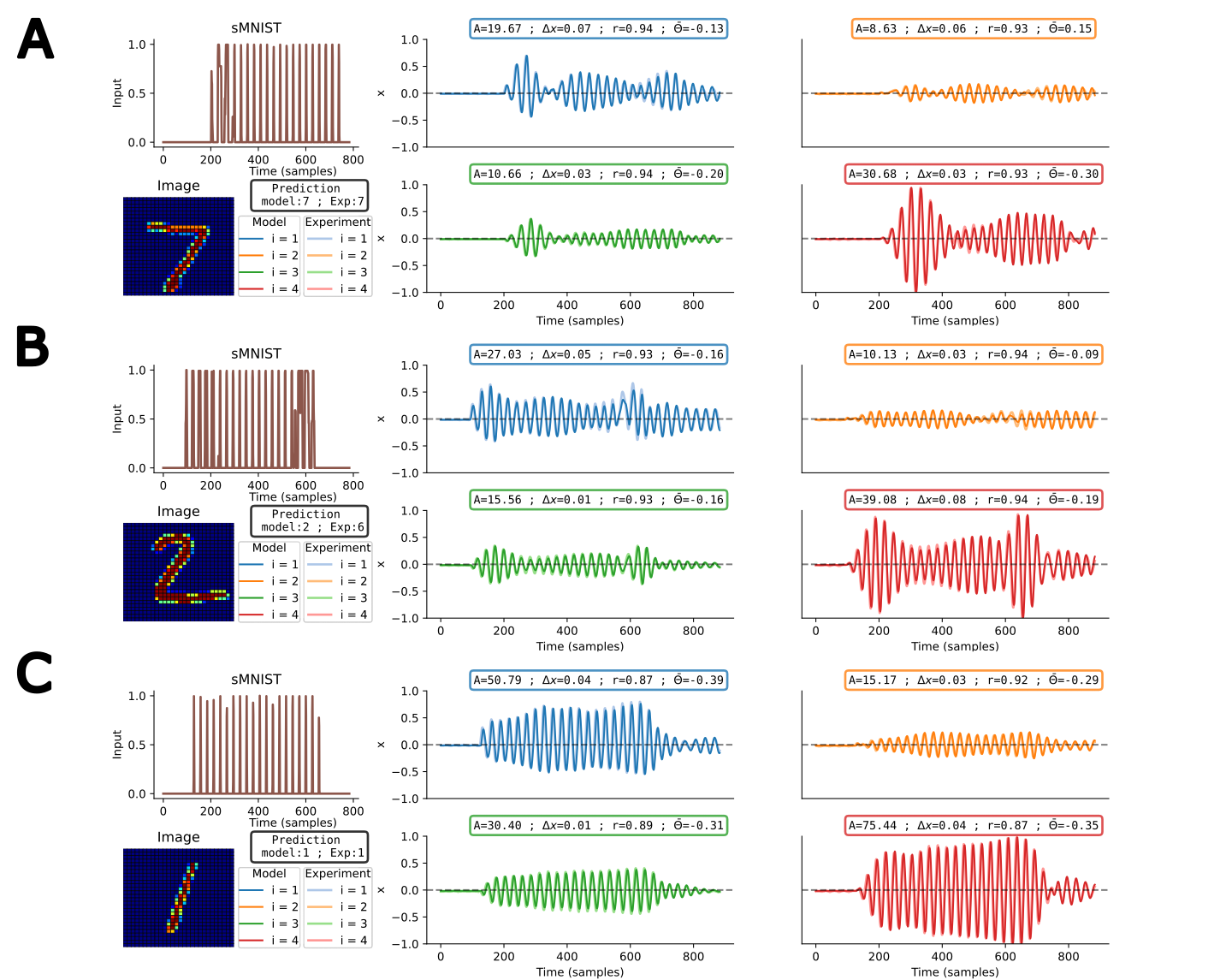}
    \caption{Representative dynamics samples from runs of velocity-coupled
    network.
    Left column displays input in both sMNIST (serialized) and MNIST (image)
    formats.
    Center and right columns show the corresponding node dynamics, with light
    traces representing analog implementation and dark traces representing
    digital twin dynamics.}
    \label{fig:Y-Net_samples}
\end{figure*}

Once implemented, we can visualize some sample network runs
(Fig.~\ref{fig:Y-Net_samples}).
We then proceed with the analysis as in the main text.
First, we look at the performance of the network
(Fig.~\ref{fig:Y-Net_performance}), similar to the amplitude-coupled case, we
find that when used as a reservoir, the network performs better.
However, the performance of the velocity-coupled network is better than that of
the amplitude-coupled network.
A Venn diagram was computed using the same procedure as in
Fig.~\ref{fig:apx_venn}.
The velocity-coupled network achieves a better reproduction of the label of its
digital twin ($58.9\%$, Fig.~\ref{fig:Y-Net_Venn}).
The error metrics analysis (Fig.~\ref{fig:Y-Net_errors}) shows that the
velocity-coupled network exhibits smaller errors than the amplitude-coupled
network (Fig.~\ref{fig:4}~D).
Finally, the velocity-coupled network was used as a reservoir with an SVM
trained to decode the network dynamics
(Fig.~\ref{fig:Y-Net_SVM_timeseries}). The SVM was trained using the variables $x$ and $y$, so that the readout has the same dimensionality as the Hilbert
decoder (`Exp. full' and `Model full' curves in
Fig.~\ref{fig:Y-Net_SVM_timeseries}). The SVM was also trained using only the
$x$ variable (`Exp.' and `Model' curves in Fig.~\ref{fig:Y-Net_SVM_timeseries})
for comparison with the amplitude-coupled network.
Similarly to the amplitude-coupled network, the SVM readout enables the analog and
digital networks to achieve comparable performance
(Fig.~\ref{fig:Y-Net_SVM_timeseries}), demonstrating that the analog
implementation preserves the information of its digital twin.
\begin{figure}
    \centering
    \includegraphics[width=\linewidth,angle=270]{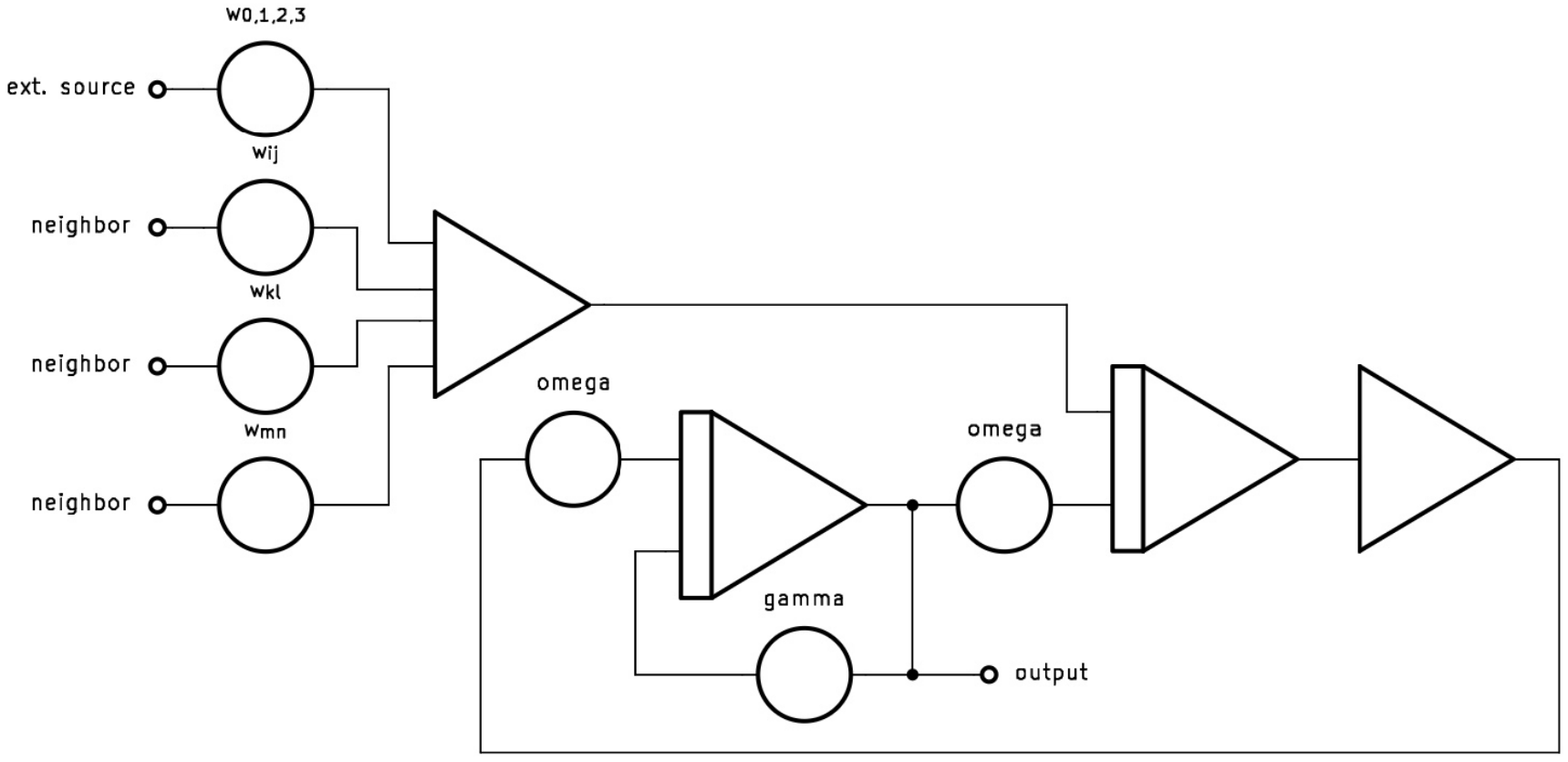}%Y-Net_circuit.jpeg}
    \caption{Circuit diagram for the oscillatory unit of the velocity coupled network.}
    \label{fig:Y-Net_Circuit}
\end{figure}

\end{document}